\documentclass[twocolumn,aps,epsfig,nofootinbib,preprintnumbers]{revtex4}

\usepackage{graphicx}
\usepackage{epstopdf}
\usepackage{latexsym}
\usepackage{amssymb}
\usepackage{amsmath}
\usepackage{color}
\usepackage{float}
\usepackage{mathrsfs}
\usepackage{multirow}
\usepackage{tabularx,booktabs}
\newcolumntype{Y}{>{\centering\arraybackslash}X}

\usepackage[center]{subfigure}
\usepackage{makecell}
\setcellgapes{4pt}
\usepackage[normalem]{ulem}

 \newcommand{\bq}{\begin{equation}}
 \newcommand{\eq}{\end{equation}}
 \newcommand{\bqn}{\begin{eqnarray}}
 \newcommand{\eqn}{\end{eqnarray}}
 \newcommand{\nb}{\nonumber}
 \newcommand{\lb}{\label}
 
\begin{document}

\title{Gravitational waveforms and radiation  powers of the triple system PSR J0337+1715 in modified theories of gravity}

 \author{Xiang Zhao$^{1,2}$}
%
\author{Chao Zhang$^{1,2}$}
 %
 \author{Kai Lin$^{3, 4}$}
%
\author{Tan Liu$^{5,6}$}
%
\author{Rui Niu$^{5,6}$}
%
\author{Bin Wang$^{7, 8}$}
%
\author{Shaojun Zhang$^2$}
%
\author{Xing Zhang$^{5,6}$} 
%
\author{Wen Zhao$^{5,6}$}
%
\author{Tao Zhu$^2$}
%
%
\author{Anzhong Wang$^{1, 2}$\footnote{ Corresponding Author}}
\email{Anzhong_Wang@baylor.edu}

\affiliation{$^1$ GCAP-CASPER, Physics Department, Baylor University, Waco, TX 76798-7316, USA}
\affiliation{$^2$ {Institute for Theoretical Physics $\&$ Cosmology, Zhejiang University of Technology,} Hangzhou 310032, China}
\affiliation{$^3$ Hubei Subsurface Multi-scale Imaging Key Laboratory, Institute of Geophysics and Geomatics, China University of Geosciences, Wuhan, Hubei, 430074, China}
\affiliation{$^4$ Escola de Engenharia de Lorena, Universidade de S\~ao Paulo, 12602-810, Lorena, SP, Brazil}
 \affiliation{$^5$ CAS Key Laboratory for Researches in Galaxies and Cosmology, Department of Astronomy, \\
University of Science and Technology of China, Chinese Academy of Sciences, Hefei, Anhui 230026, China}
\affiliation{$^6$ School of Astronomy and Space Science, University of Science and Technology of China, Hefei 230026, China}
\affiliation{$^7$ Center for Gravitation and Cosmology, Yangzhou University, Yangzhou 225009, China}
\affiliation{$^{8}$ School of Aeronautics and Astronautics, Shanghai Jiao Tong University, Shanghai 200240, China}

\date{\today}

\begin{abstract}

In this  paper, we study the gravitational waveforms, polarizations  and radiation powers of the first relativistic triple systems PSR J0337 + 1715, observed in 2014, 
by using the post-Newtonian approximations to their lowest order. Although they cannot be observed either by current or next generation of the detectors, they do
provide useful information to test different theories of gravity. In particular, we carry out the studies in three different theories, General Relativity, Einstein-{\ae}ther theory  and Brans-Dicke gravity. The tensor modes $h_{+}$ and $h_{\times}$ exist in all three theories and have almost equal amplitudes. Their frequencies 
are all peaked at two locations,  $ f^{+, \times}_1 = 0.068658 \mu$Hz and $ f^{+, \times}_2 =14.212 \mu$Hz, which are  about twice the outer and 
inner orbital frequencies of the triple system, as predicted in GR.  In $\ae$-theory, all the six polarization modes are different from zero, but the breathing ($h_b$) and 
longitudinal ($h_L$) modes are not independent and also peaked at two  frequencies. A somehow surprising result is that, for $h_{b}$ and $h_{L}$, the peaked 
frequencies are not twice the outer and inner orbital frequencies, as for the $h_{+}$ and $h_{\times}$ modes, but instead,  they are almost equal to them, 
$ f^{b, L}_{1} = 0.045772 \mu$Hz and $f^{b, L}_{ 2} = 7.0947 \mu$Hz. A similar phenomenon is also observed  in BD gravity, in which only 
the three modes $h_{+},\; h_{\times}$ and $h_{b}$ exist, where $ f^{b}_{ 1}$ and $ f^{b}_{2}$ are almost equal to outer and inner orbital frequencies. We also 
study the radiation powers, and find that the  quadrupole emission  in each of the three theories has almost the same amplitude, but the dipole emission can be as big 
as the quadrupole emission  in $\ae$-theory. This provides a very promising window to obtain severe constraints on $\ae$-theory by the multi-band gravitational wave 
astronomy. 

\end{abstract}

\maketitle

\section{Introduction}
\renewcommand{\theequation}{1.\arabic{equation}}
\setcounter{equation}{0}

The concept of gravitational waves (GWs) was first developed by  Einstein in 1916 right  following his general theory of relativity (GR). He proposed that GWs are the ripples of spacetimes that are propagating with the speed of light \cite{Einstein}. A century passed, the Laser Interferometer Gravitational-Wave Observatory (LIGO) first verified his GW theory by directly detecting the GW signal on Sep. 14,  2015 \cite{GW150914}. After this, ten more GWs have been detected by LIGO Scientific Collaboration and Virgo Collaboration \cite{GW151226, GW170104, GW170608, GW170814, GW170817,GWs}. The sources of these eleven GW events are all binary systems of black holes, except for the event GW170817, which is a binary system of  neutron stars  \cite{GW170817}. 

\begin{figure}[h!]
\includegraphics[width=\linewidth]{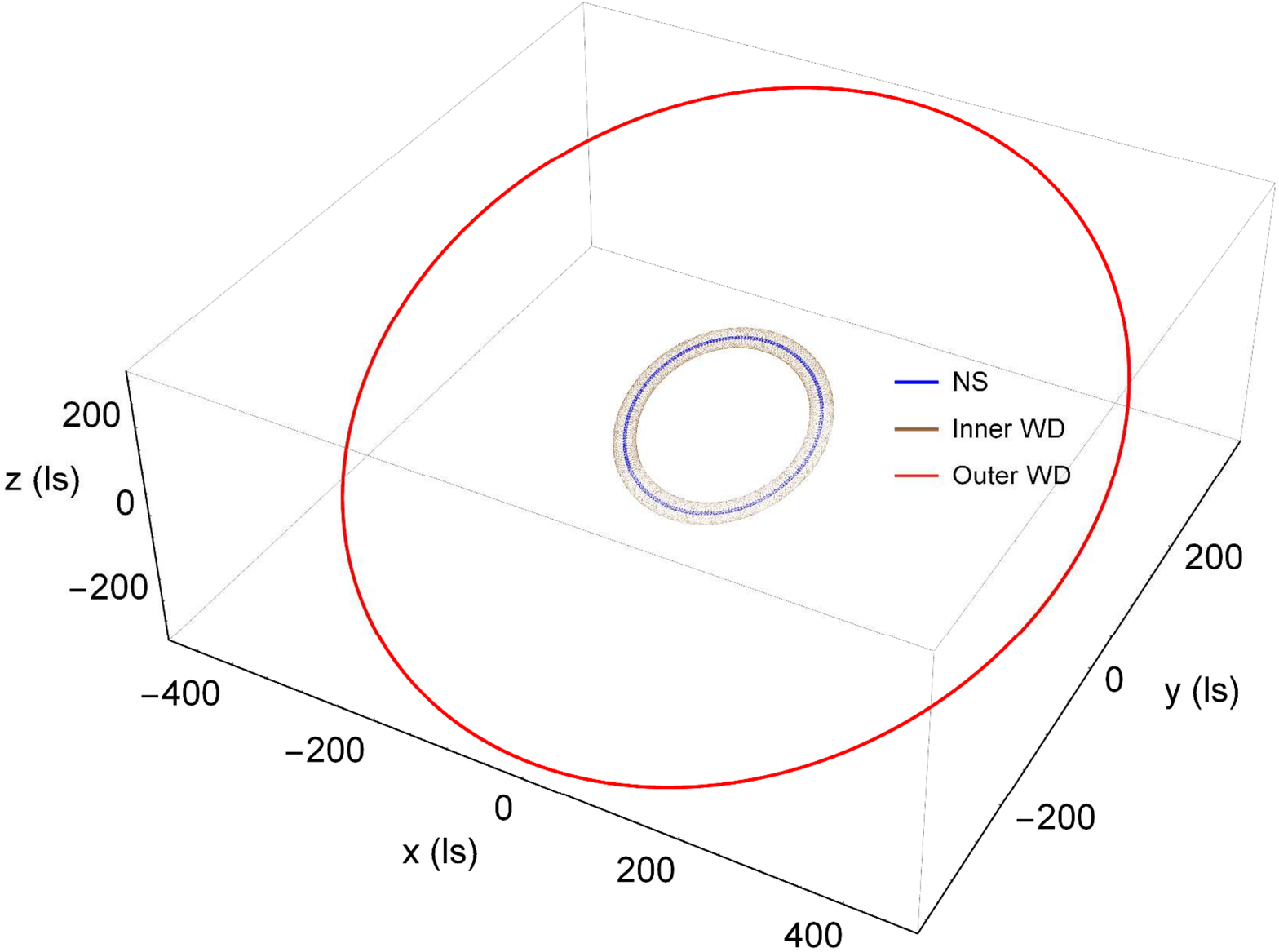} \\
\includegraphics[width=\linewidth]{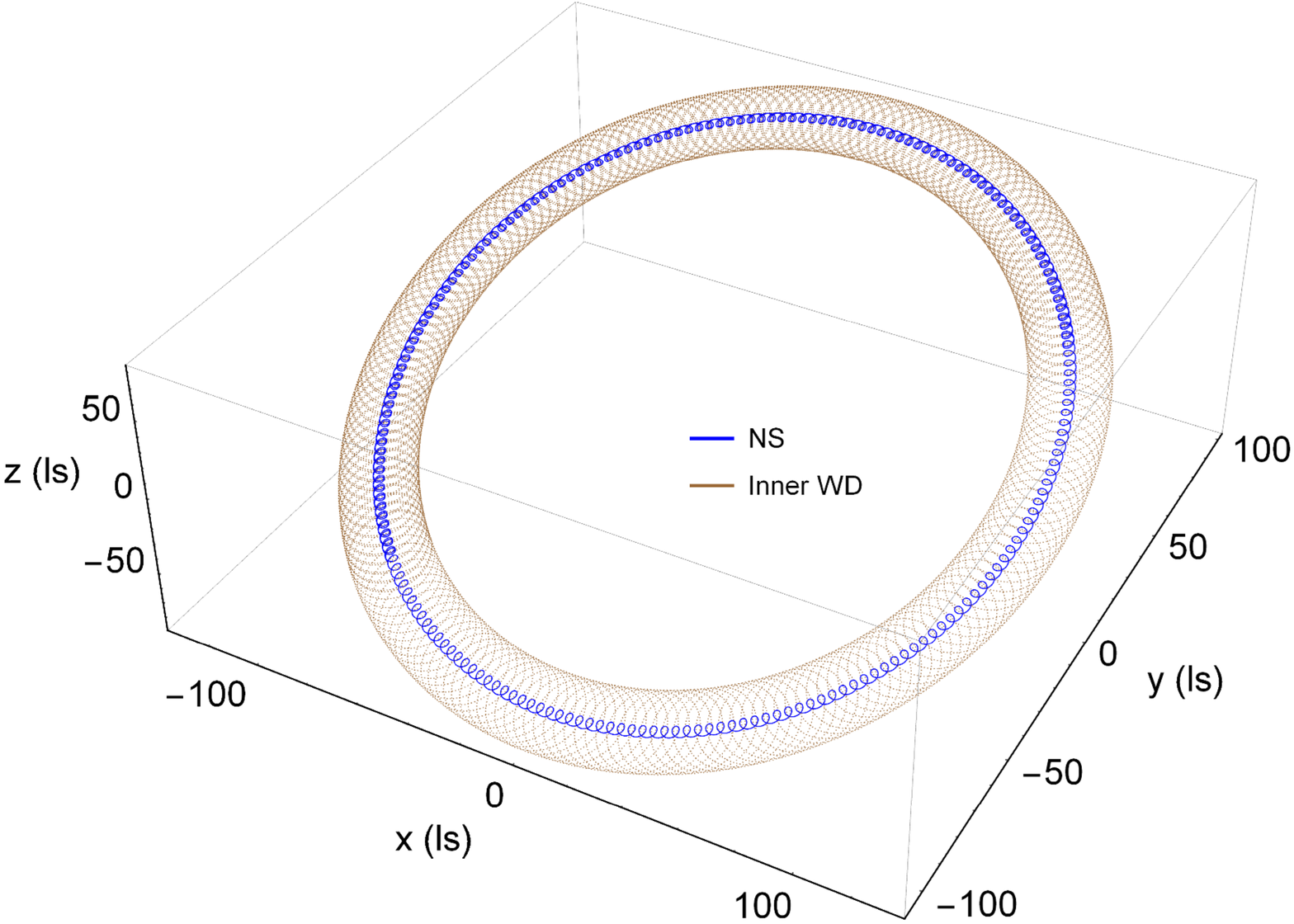} 
\caption{
Orbit of the NS, inner WD and the outer WD  where ls stands for the light second. This plot shows the trajectories observed in the center-of-mass coordinate system for about 330 days starting from 01-04-2012 \cite{Shao16}.      
} 
\label{fig1}
\end{figure}

In fact, there are about $13\%$ of low-mass stellar systems containing three or  more stars \cite{FC17}, and $96\%$ of low-mass binaries with periods shorter than 3 days which are part of a larger hierarchy \cite{Tok06}. Recently, a realistic triple system   was observed, named as PSR J0337 + 1715 (J0337) \cite{Ransom14}, which consists of an inner  binary and a third companion. The inner binary consists of a pulsar  with mass  $m_1 = 1.44 M_{\bigodot}$ and a white dwarf (WD) with mass  $m_2 = 0.20 M_{\bigodot}$ in a 1.6 day orbit.  The outer  binary consists of the inner binary and a second dwarf with mass $m_3 = 0.41 M_{\bigodot}$ in a  327 day orbit.  The two orbits are very circular with its eccentricities $e_I \simeq 6.9\times 10^{-4}$ for the inner binary and $e_O \simeq 3.5  \times 10^{-2}$ for the outer orbit. The two orbital planes  are remarkably  coplanar with an inclination  $\lesssim 0.01^{o}$ [See Fig. \ref{fig1}].
 
A triple system is an ideal place to test the strong equivalence principle \cite{Shao16}. Remarkably, after 6-year observations, recently  it  was found that the accelerations of the pulsar and its nearby white-dwarf companion differ fractionally by no more than $2.6 \times 10^{-6}$ \cite{Archibald18}, which provides the most severe constraint on the violation of the strong equivalence principle. 

In this paper, we investigate this triple system in three different theories of gravity, General Relativity (GR), Einstein-aether theory (${\ae}$-theory) and Brans-Dicke (BD) gravity, by using the post-Newtonian approximations to their lowest order. We shall pay particular attention on the differences predicted by these theories. Although neither the current generation of detectors nor the next one can detect the GWs emitted by this system \footnote{The frequency of the GWs emitted by this system is about $10^{-8}$ Hz. With this frequency, only the pulsar timing array (PTA),  such as IPTA and SKA, can detect such GWs. But, the amplitudes of these GWs are far below the sensitivities of these detectors.}, it serves well as a realistic example to show clearly the different predictions from each of these theories. In particular, we shall study, gravitational waveforms, their polarizations, Discrete Fourier transfom (DFT) of polarizations as well as the radiation powers. Among the modified theories of gravity, ${\ae}$-theory locally breaks the Lorentz symmetry by introducing a globally time-like unit vector field (the ${\ae}$ther) \cite{Jacobson}, while in BD gravity the gravitational interaction is mediated by   both a scalar and a tensor fields \cite{Brans}. 

Specifically, the paper is organized as follows: In Sec. II, we study the gravitational waveforms, polarization modes and their DFTs in GR, $\ae$-theory, and BD gravity respectively, while in Sec. III we investigate the radiation powers of the GW in each theory. Due to the presence of the extra vector and scalar fields in $\ae$-theory, and the extra scalar  filed {in} BD gravity,  the total emission power is different in each of the three theories. In particular, we find that the dipole emission in $\ae$-theory can be as large as the quadrupole emission in GR, which can provide a very promising window to obtain severe constraints on $\ae$-theory by the multi-band gravitational wave astronomy \cite{AS16}. There is also an appendix, in which we present a very brief introduction to ${\ae}$-theory \cite{Jacobson}.

Before turning to the next section, we would like to note that,  in the framework of \ae-theory, Foster \cite{Foster07} and Yagi {\em et al} \cite{Yagi14} derived the metric and equations of motion to the 1PN order for a N-body system. Recently, Will applied them to study the 3-body problem and obtained the accelerations of a 2-body system in the presence of the third body at the quasi-Newtonian order \cite{Will18}. For nearly circular coplanar orbits, he also calculated the 
  ``strong-field" Nordtvedt parameter $\hat{\eta}_{N}$. For triple system J0337, ignoring the sensitivities of the  two white-dwarf companions,  Will found that  $\hat{\eta}_{N}$   is given by $\hat{\eta}_{N} = s_{1}/(1-s_1)$, where $s_1$ denotes the sensitivity of the pulsar. 

In this paper,  we will adopt the following conventions: All the repeated indices $i, j, k, l \; (i, j, k, l = 1, 2, 3)$ will be summed over regardless of their vertical poition, while repeated indices $a, b, c \; (a, b, c = 1, 2, 3)$ will not be summed over unless the summation is explicitly indicated. We set the speed of light equal to one ($c=1$). The metric signature is $(-, +, +, +)$. Since J0337 is a triple system, there exists no analytical orbit, we are using the numerical orbit supplied by Dr. Lijing Shao \cite{Shao16}.

\begin{figure}[h!]
\includegraphics[width=\linewidth]{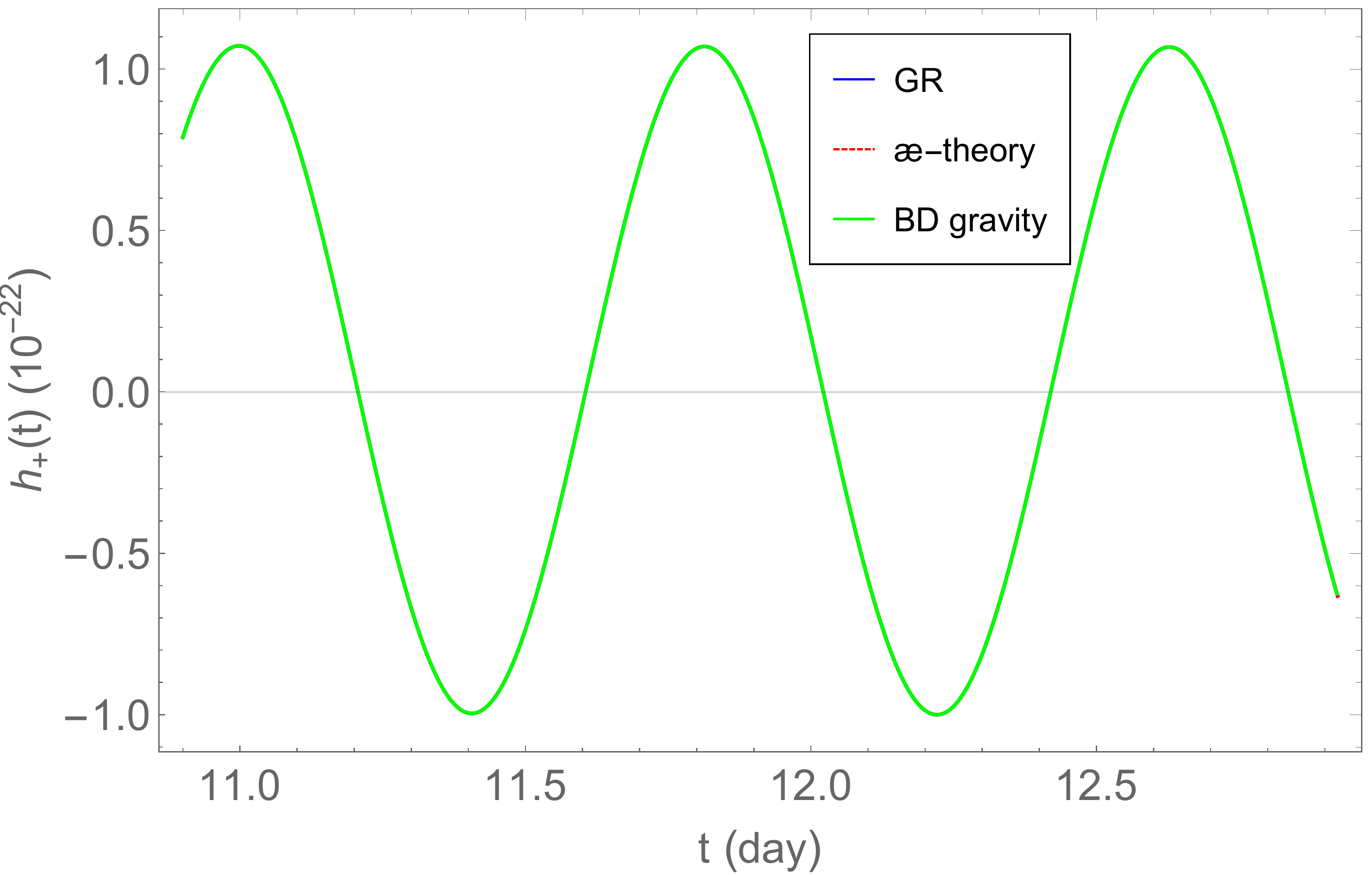} \\
\includegraphics[width=\linewidth]{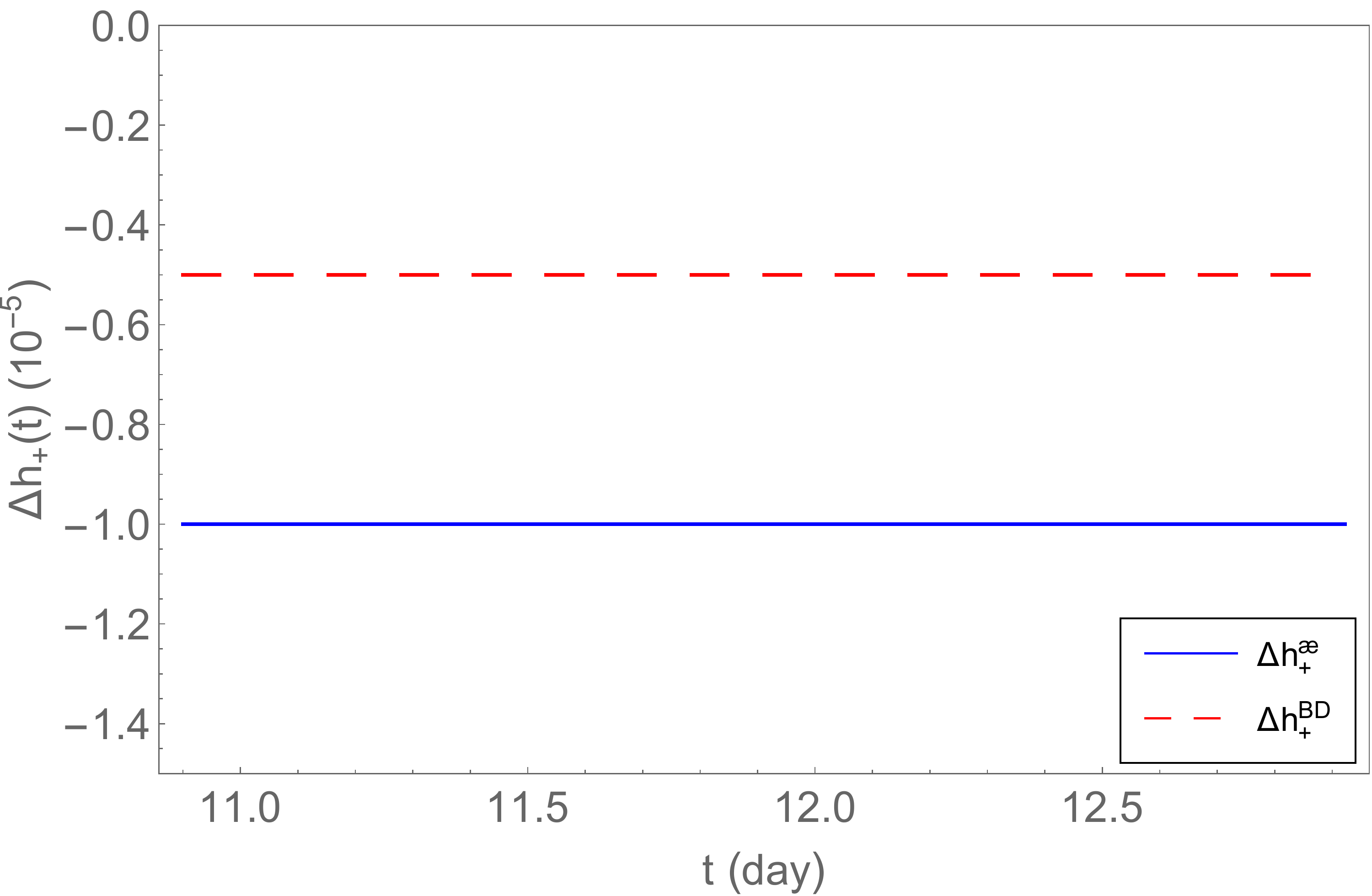} 
\caption{In the upper panel, the plus polarization $h_{+}$, respectively,  in GR, {\ae}-theory and BD gravity are plotted, while  their relative differences with respect to GR, given by
 Eqs.(\ref{4.1db}) and (\ref{3.11b}) are plotted out in the bottom panel. }  
\label{fig2}
\end{figure}

\begin{figure}[h!]
\includegraphics[width=\linewidth]{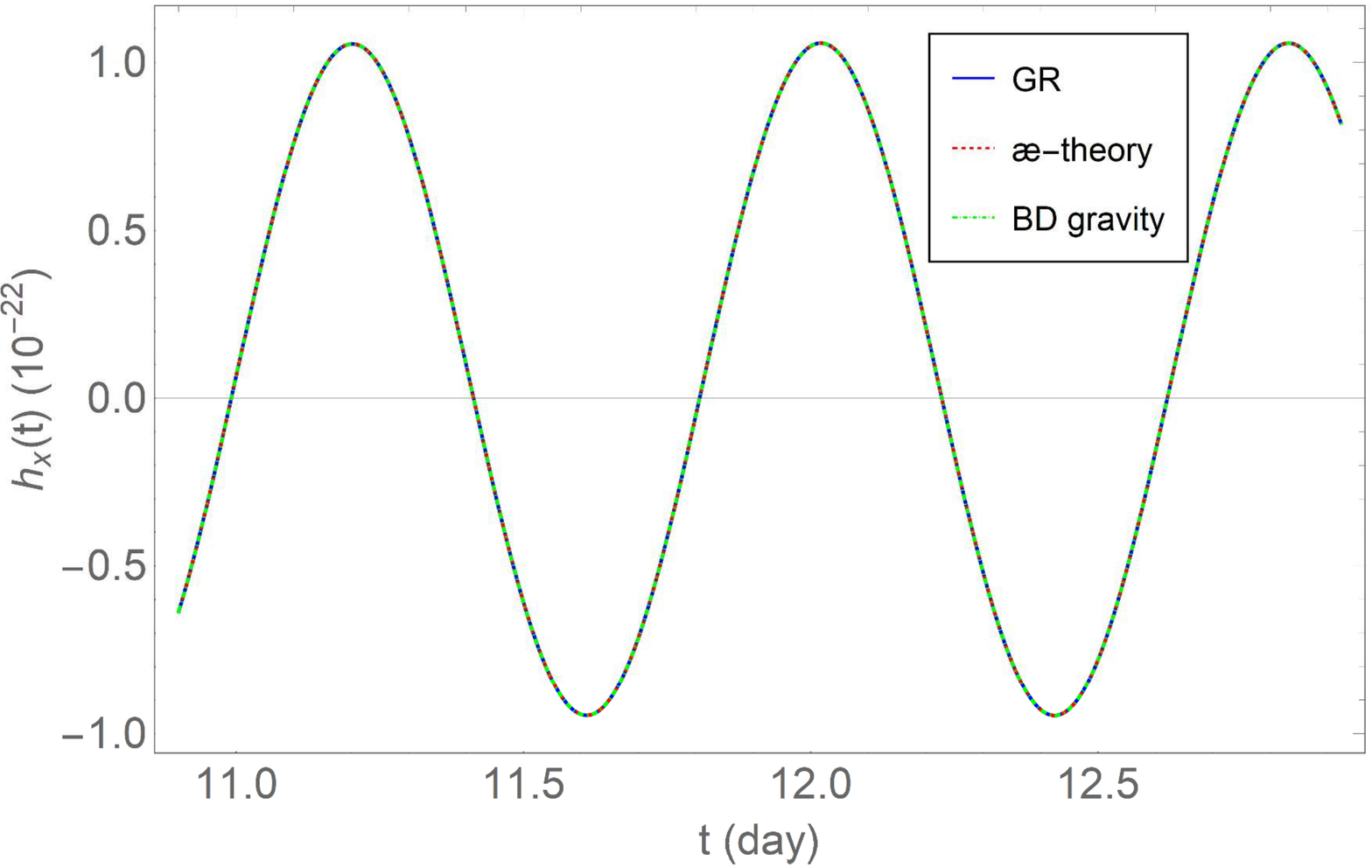} \\
\includegraphics[width=\linewidth]{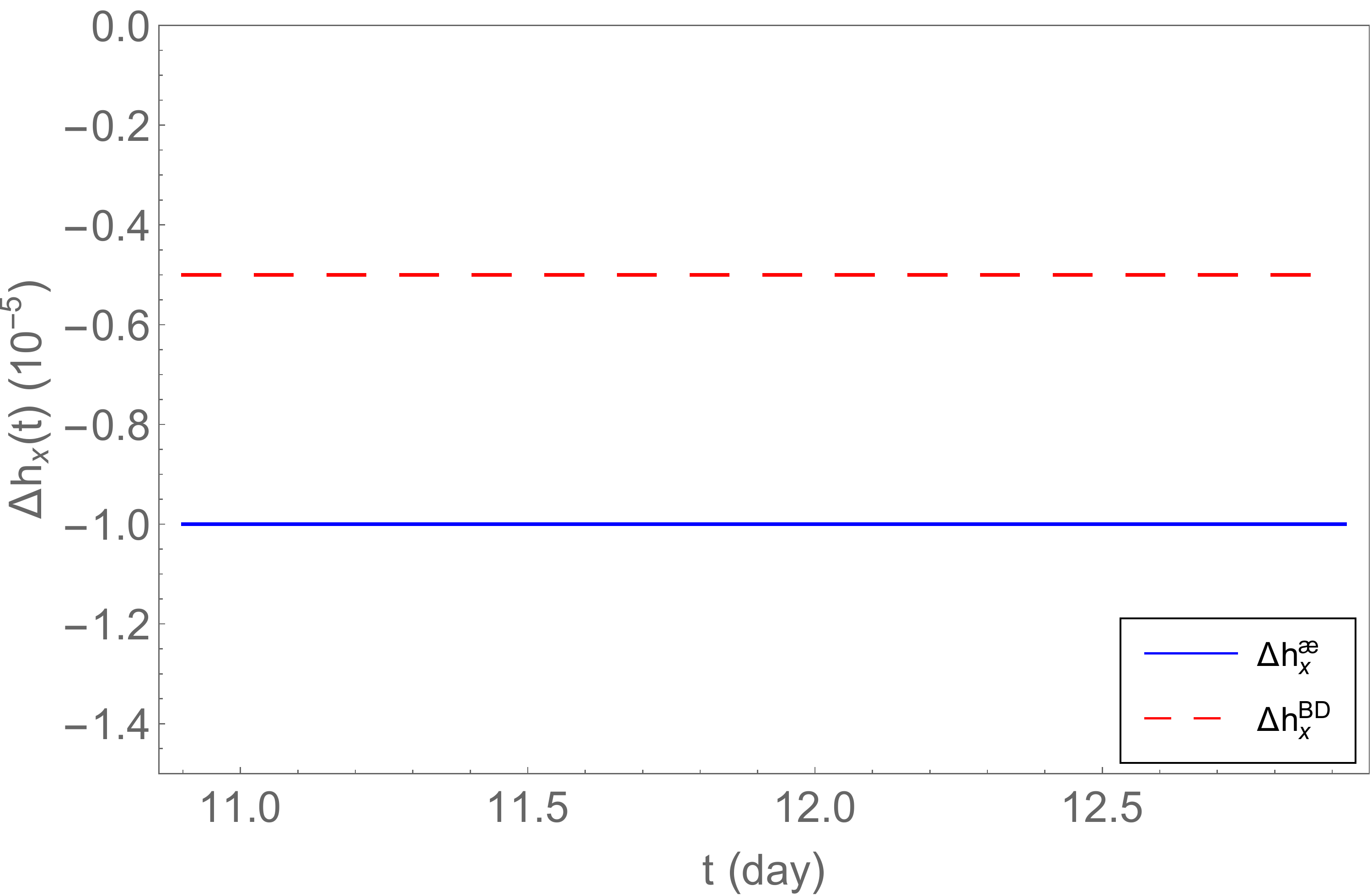} 
\caption{  In the upper panel, the cross polarization $h_{\times}$, respectively,  in GR, {\ae}-theory and BD gravity are plotted, while  their relative differences with respect to GR, given by
 Eqs.(\ref{4.1db}) and (\ref{3.11b}) are plotted out in the bottom panel. } 
\label{fig3}
\end{figure}

\section{Gravitational Waveforms and their Polarizations}
\renewcommand{\theequation}{2.\arabic{equation}}
\setcounter{equation}{0}

When a GW passes  two test masses, the distance between them will be changed. Assuming that   $\zeta^i$ denotes  the spatial coordinates between these two test masses in the Minkowski coordinates $x^{\mu} = (t, x, y, z)$, 
the equations of the geodesic deviation in the weak-field approximations read, 
\bq
\lb{3.1}
\ddot{\zeta}_i=-R_{0i0j}\zeta^j,
\eq
where $R_{\mu\nu\alpha\beta}$ denotes the linearized Riemann tensor, which is determined by the field equations of a given theory. Different theory yield different components of $R_{0i0j}$. Therefore, in the following we shall consider GR, \ae-theory and  BD gravity separately.   

\subsection{Gravitational Waveforms and Their Polarizations in GR}

In GR, the equations of the geodesic deviation  take the form \cite{MM}, 
\bq
\lb{3.1}
\ddot{\zeta}_i=-R_{0i0j}\zeta^j = \frac{1}{2}\ddot{h}_{ij}^{TT}\zeta^j,
\eq
where
\bq
\lb{3.2}
h_{ij}^{TT}(t,{\bf x}) = \frac{2 G_N}{R} \ddot{Q}_{ij}^{TT}(t - R),
\eq
with $R \equiv |{\bf x}|$ denoting the distance from the observer to the source  and $G_N$ denoting the Newtonian constant.  
In the above equations, $TT$ represents the transverse-traceless part of the tensor, which can be obtained by applying $TT$ operator $\Lambda_{ijkl}$,
\bqn
\lb{3.3}
\Lambda_{ijkl}({\bf N})  = P_{ik} P_{jl} - \frac{1}{2} P_{ij} P_{kl},
\eqn
where $P_{ij}({\bf N}) = \delta_{ij} - N_i N_j$. Let's introduce a set of orthorgonal bases, (${\bf e}_X, {\bf e}_Y, {\bf e}_Z$), where ${\bf e}_Z \equiv \bf{N}$ denotes the propagation direction of the GW from the source to the observers.
Thus, (${\bf e}_X, {\bf e}_Y$) forms a plane orthogonal to the propagation direction $\bf{N}$ of the GW. In the $(t, x, y, z)$ coordinates, they are given by,
\bqn
\lb{3.4}
{\bf e}_{X} &=& \left(\cos\vartheta\cos\varphi, \cos\vartheta\sin\varphi, - \sin\vartheta\right), \nb\\
{\bf e}_{Y} &=& \left(-\sin\varphi, \cos\varphi, 0\right), \nb\\
{\bf e}_{Z} &=& \left(\sin\vartheta\cos\varphi, \sin\vartheta\sin\varphi, \cos\vartheta\right),
\eqn
where $\vartheta$ and {$\varphi$} are the two spherical angular coordinates. In the case of J0337, $\vartheta$ and $ \varphi$ are   0$^\circ$ and 270$^\circ$,  respectively \cite{Ransom14}.
In GR, $h_{ij}^{TT}$  has only two degrees of freedom corresponding to the  plus (``+") and cross (``$\times$") polarizations, which are given by, 
\bqn
\lb{3.5}
h_+^{GR} = \frac{G_N}{R}\ddot{Q}_{kl} e_+^{kl}, \;\;
h_\times^{GR} =  \frac{G_N}{R}\ddot{Q}_{kl} e_\times^{kl},
\eqn
where $e_+^{kl} \equiv e_X^k e_X^l-e_Y^k e_Y^l$, $e_\times^{kl} \equiv e_X^k e_Y^l+e_Y^k e_X^l$.   In Figs. \ref{fig2} and \ref{fig3}, we plot these two polarizations, from figures it can be seen that the amplitudes of both polarizations are about $10^{-22}$, which is in the range of the designed sensitivity of the current generation of the ground-based detectors,  such as, LIGO, Virgo and {KAGRA}, but not their frequencies (See, e.g. \cite{Moore15}). This is because the orbital frequency of J0337 is out of the observational bands of  the current detectors. You can find the frequencies of these two polarizations easily from Figs. \ref{fig4} and \ref{fig5}.

To obtain the Fourier transform for each polarization mode, instead of using the continuous Fourier transform,  
$\tilde{h}_A(f) = (2\pi)^{-1} \int{h_A(t) e^{-i2\pi f t} dt}$, we adopt the discrete Fourier transform (DFT). First, the time interval of signal $h_A(t)$ is divided into $N-1$ subintervals, $t= \left\{t_0, t_1, ..., t_{N-1}\right\}$, $t_m= m \Delta t$, $\Delta t$ is the length of subinterval.  Then we have a list of dicrete values $h_A^m \equiv h_A(t_m),$ and we get the DFT accodrding to the following formula
\bq
\lb{3.5a}
\tilde{h}_A^n = {\frac{1}{N} \sum_{m = 0}^{N-1}h_A^m e^{-i2\pi n m/N}},
\eq
where $ \tilde{h}_A^n \approx \tilde{h}_A(f)$, $f= \left\{f_0, f_1, ..., f_{N-1}\right\}$, $f_n = n/(N\Delta t)$. In this paper, we shall use the built-in function of software \textbf{Mathematica} to calculate DFT of waves. With this in mind,  in Figs. \ref{fig4} and \ref{fig5} we plot $\tilde{h}_+^{GR}(f)$ and $\tilde{h}_{\times}^{GR}(f)$, respectively. In these figures, there are two peaks and the corresponding frequencies are 
 \bqn
 \lb{3.5b}
 f^{+, \times}_1 &=& 0.068658 \mu \mbox{Hz} \simeq  2.0 f_o,\nb\\
 f^{+, \times}_2 &=& 14.212 \mu \mbox{Hz} \simeq 1.9 f_i,
 \eqn
where 
$f_o$ and $ f_i$ represent outer and inner orbital frequencies of the triple system J0337 \cite{Ransom14},
\bqn
\lb{3.5c}
 f_o = 0.035 \mu \mbox{Hz},\;\;\;  
 f_i =  7.103 \mu \mbox{Hz}. 
 \eqn
Thus, $f^{+, \times}_1$ and $f^{+, \times}_2$ are about twice of the outer and inner orbital frequencies of the triple system, but not exactly. In GR, for a binary system  the GW frequency is exactly equal to two times of their orbital frequency  \cite{MM}.
However, it must be noted that here the difference is due to the presence of the third component of the triple system.  
 
\begin{figure}[h!]
\includegraphics[width=\linewidth]{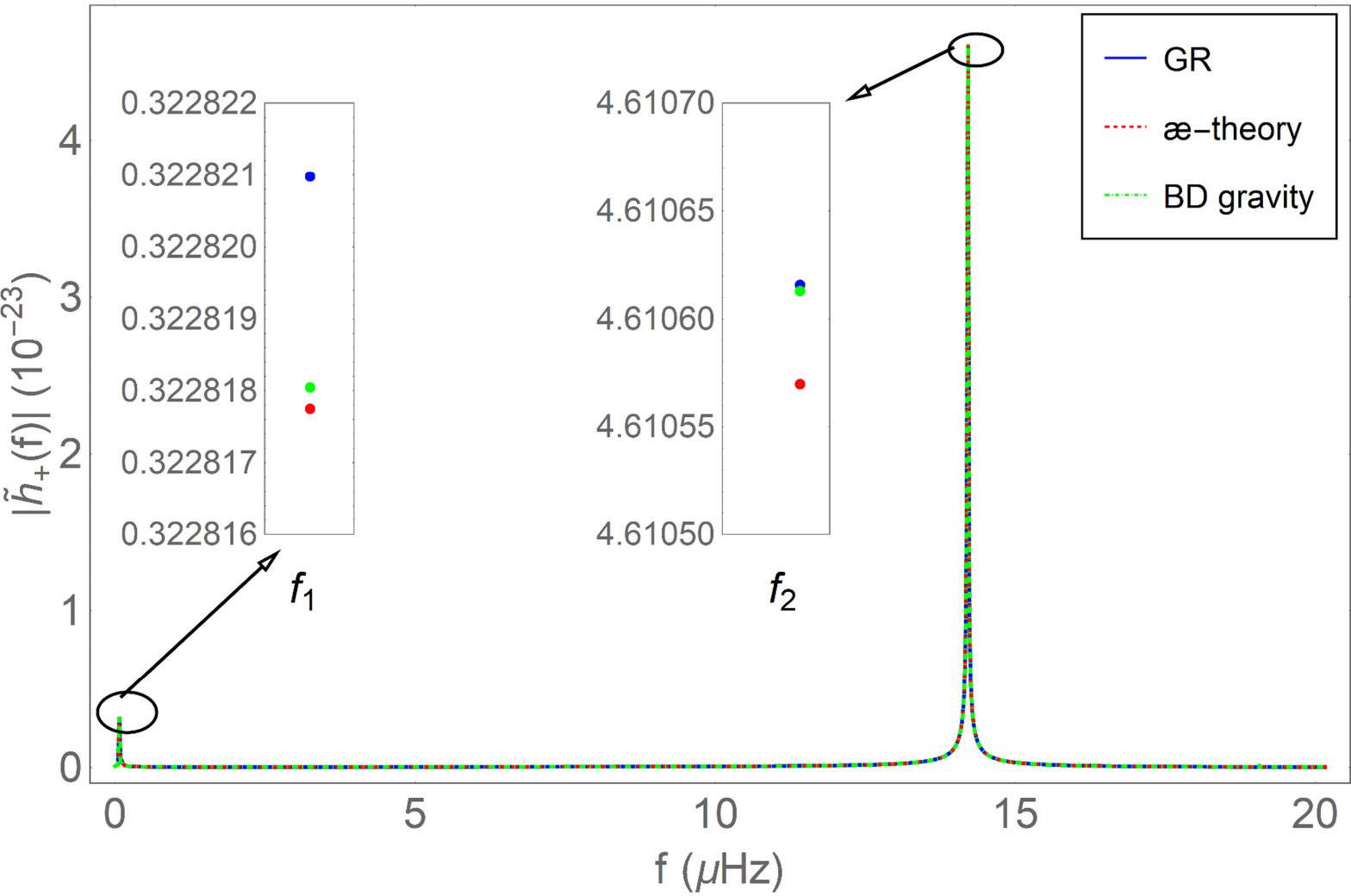} \\
\includegraphics[width=\linewidth]{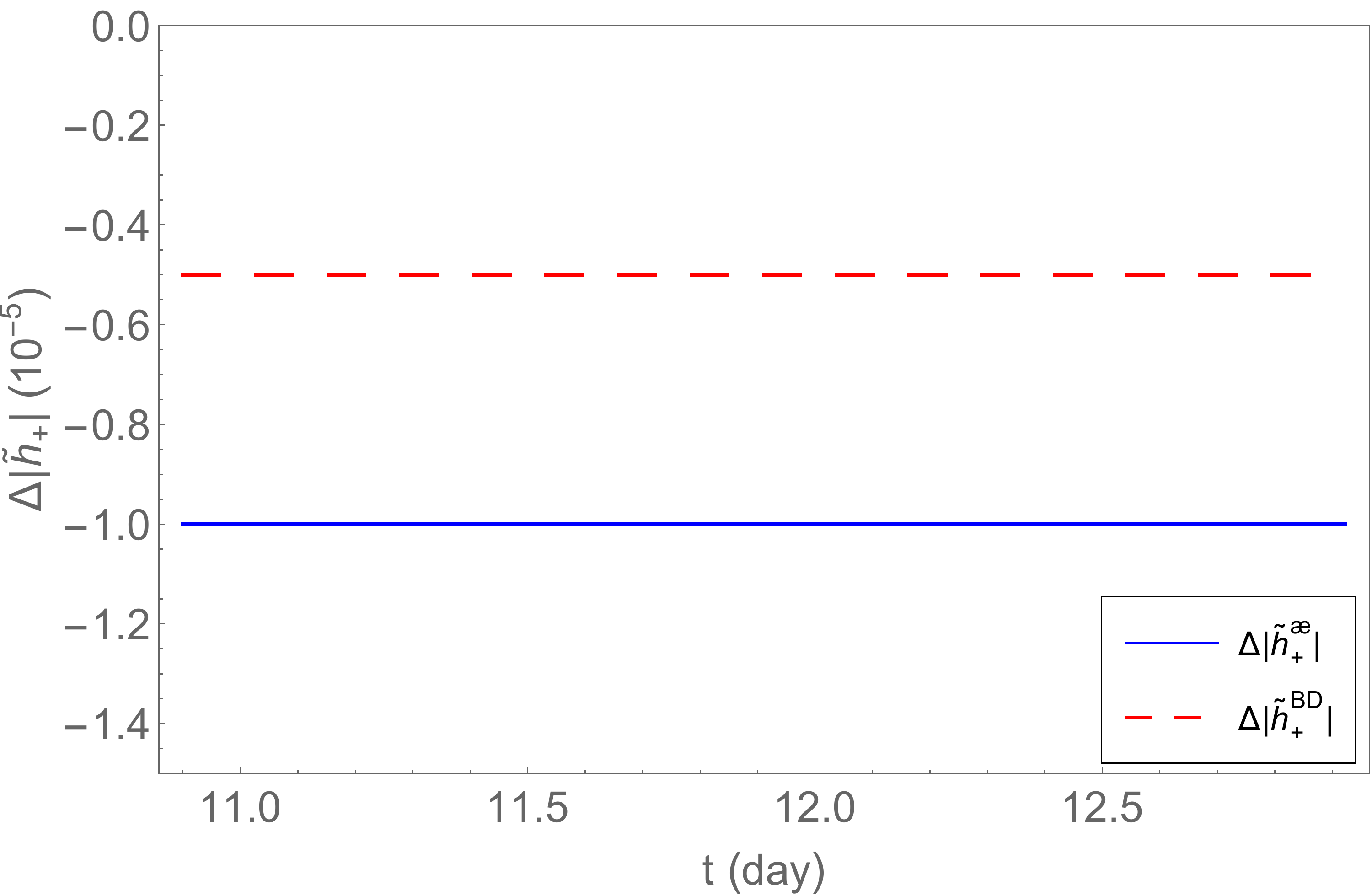} 
\caption{In the upper panel, DFTs of the plus polarization $\tilde{h}_+(f)$, respectively,  in GR, {\ae}-theory and BD gravity, are plotted out, in which the two peaked frequencies have been marked. The inserted images show the tiny differences at the two peaked frequencies among the three different theories, where $f_1= 0.068658\mu$Hz, $f_2=14.212\mu$Hz.  In the bottom panel,    their relative differences with respect to GR, given, respectively,  by Eqs.(\ref{4.1dc}) and (\ref{3.11c}) are plotted out. } 
\lb{fig4}
\end{figure}

\begin{figure}[h!]
\includegraphics[width=\linewidth]{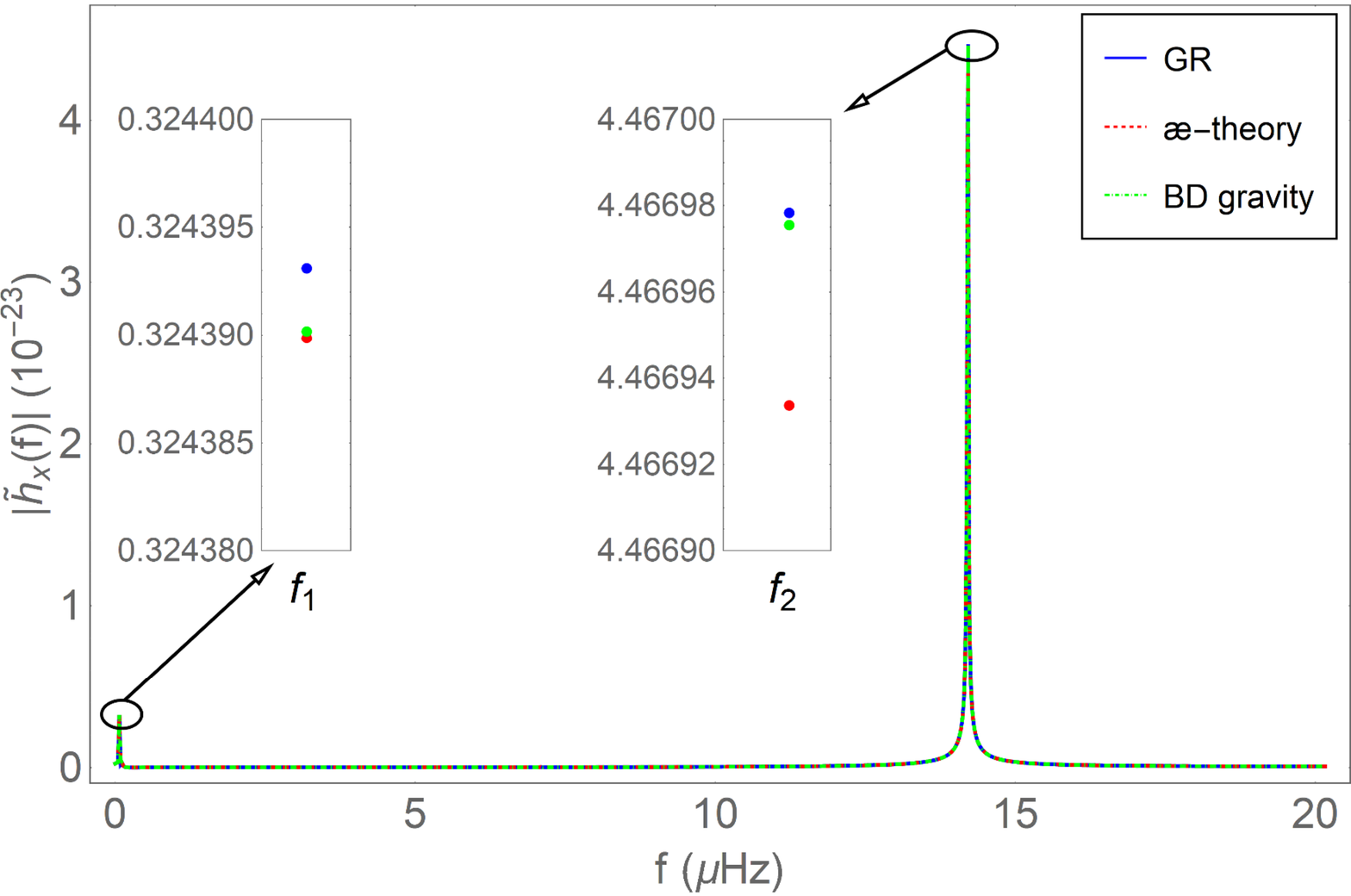} \\
\includegraphics[width=\linewidth]{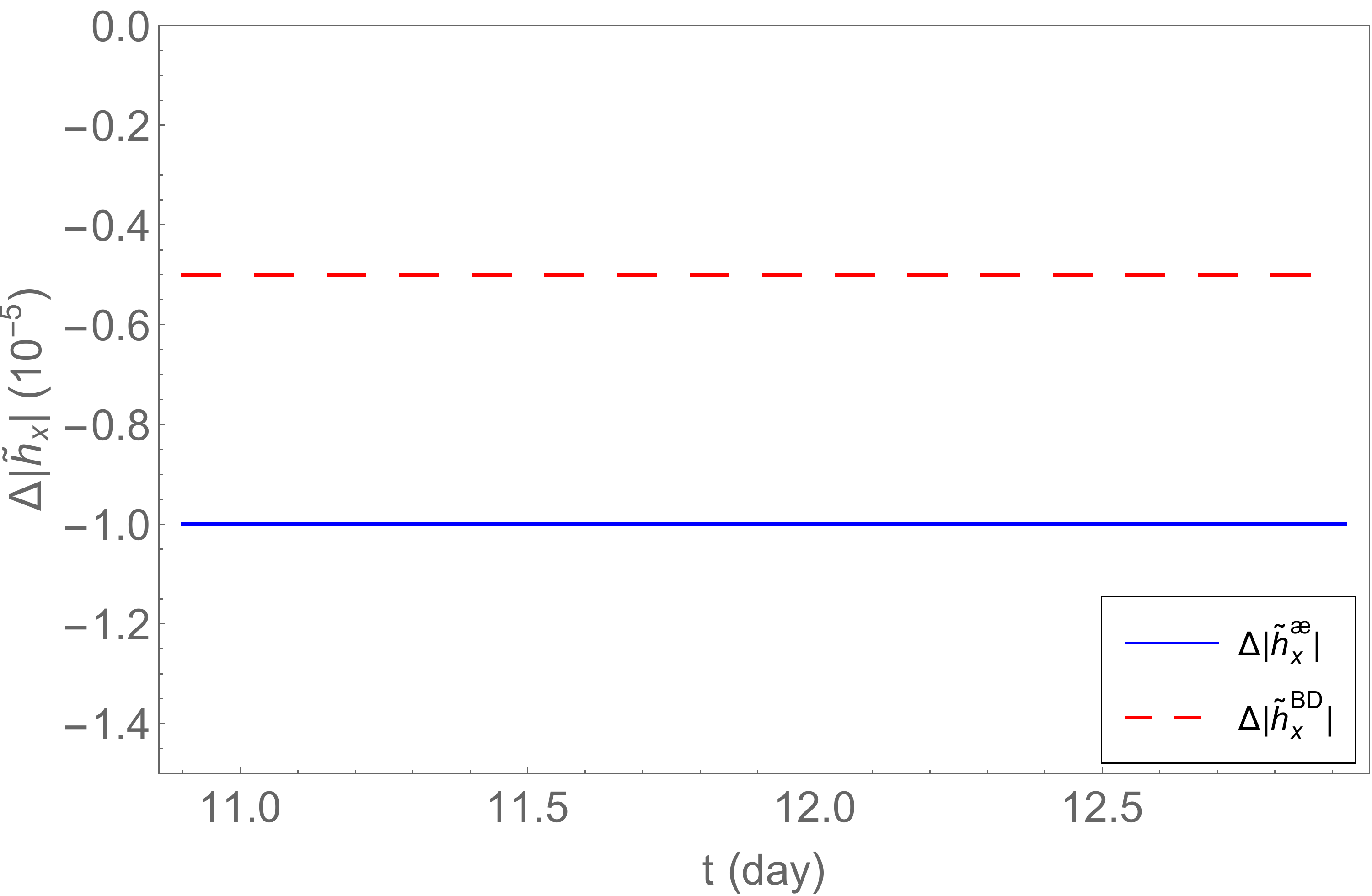} 
\caption{ In the upper panel, DFTs of the cross polarization $\tilde{h}_{\times}(f)$, respectively,  in GR, {\ae}-theory and BD gravity are plotted out, in which the two peaked frequencies have been marked. The inserted images show the tiny differences at the two peaked frequencies among the three different  theories, where $f_1= 0.068658\mu$Hz, $f_2=14.212\mu$Hz. In the bottom panel,    their relative differences with respect to GR, given, respectively,  by Eqs.(\ref{4.1dc}) and (\ref{3.11c}) are plotted out. } 
\lb{fig5}
\end{figure}

\subsection{Gravitational Waveforms and Their Polarizations in  {\ae}-theory}

In $\ae$-theory, the equation of the geodesic deviation reads \cite{Lin}, 
\bq
\lb{3.6}
\ddot{\zeta}_i=-R_{0i0j}\zeta^j\equiv\frac{1}{2}\ddot{{\cal P}}_{ij}\zeta^j,
\eq
where
\bqn
\lb{3.7} 
{\cal P}_{ij} &=& \phi_{ij}-\frac{2c_{13}}{(1-c_{13})c_{V}}\Psi^{\mathrm{I}}_{(i}N_{j)} \nb\\
              &&  -\frac{c_{14}-2c_{13}}{c_{14}(c_{13}-1)c^2_{S}}\Phi^{\mathrm{II}}N_iN_j+\delta_{ij}\Phi^{\mathrm{II}},
\eqn
and $\phi_{ij},\; \Psi^{\mathrm{I}}_{i}$ and $\Phi^{\mathrm{II}}$ are, respectively, the gauge-invariant quantities of the tensor, vector and scalar modes defined in \cite{Lin}. 
The quantities $c_V$ and $c_S$ denote the speeds of the vector and scalar modes, respectively. Due to their presence,  the polarizations of a GW have five independent modes which are given respectively by \cite{Lin}
\bqn
\lb{3.8}
h_+^{\ae} &\equiv& \frac{1}{2}\left({\cal{P}}_{XX}-{\cal{P}}_{YY}\right) = \frac{G_{\ae}}{R}\ddot{Q}_{kl} e_+^{kl},\\
 h_\times^{\ae}  &\equiv& \frac{1}{2}\left({\cal{P}}_{XY}+{\cal{P}}_{YX}\right) = \frac{G_{\ae}}{R}\ddot{Q}_{kl} e_\times^{kl}, \\
h_b^{\ae} &\equiv& \frac{1}{2}\left({\cal{P}}_{XX}+{\cal{P}}_{YY}\right)  \nb\\
&=& \frac{c_{14}G_{\ae}}{R(2-c_{14})}  \Bigg[3(Z-1)\ddot{Q}_{ij}e_Z^ie_Z^j   \nb\\
&& ~~~~~~~~~~~~~~~~~ -\frac{4}{c_{14}c_S}\Sigma_i e_Z^i+Z\ddot{I}\Bigg], \\
h_L^{\ae} &\equiv& {\cal{P}}_{ZZ} = \left[1-\frac{c_{14}-2c_{13}}{c_{14}(c_{13}-1)c_S^2}\right]h_b^{\ae}, \\
h_X^{\ae} &\equiv&\frac{1}{2}\left({\cal{P}}_{XZ}+{\cal{P}}_{ZX}\right) = \frac{2c_{13} G_{\ae}}{(2c_1-c_{13}c_-)c_V R} \nb\\
          &&  \times \left[\frac{c_{13}\ddot{Q}_{jk}e_Z^k}{(1-c_{13})c_V}-2\Sigma_j\right]e_X^j ,\\
h_Y^{\ae} &\equiv& \frac{1}{2}\left({\cal{P}}_{YZ}+{\cal{P}}_{ZY}\right) = \frac{2c_{13} G_{\ae}}{(2c_1-c_{13}c_-)c_V R} \nb\\
          &&  \times \left[\frac{c_{13}\ddot{Q}_{jk}e_Z^k}{(1-c_{13})c_V}-2\Sigma_j\right]e_Y^j ,
\eqn
where $c_{-} \equiv c_1 - c_3$, $G_{\ae} = G_{N}(1 - c_{14}/2)$, ${\cal{P}}_{XY} \equiv {\cal{P}}_{ij}e^{i}_X e^{j}_Y$, and so on. We have also defined, 
\bqn
\lb{4.1b}
Q_{ij} &=& I_{ij} - \frac{1}{3} \delta_{ij}I,\nb\\
I_{ij} &=& \sum\limits_{a} m_a x^i_a x^j_a,
\eqn
where $x^i_a$ is the location of the $a$-th body and $I \equiv I_{kk}$. For more details,  see \cite{Lin}. From the above expressions, in addition to the usual plus and cross polarization modes, $\ae$-theory predicts  three extra independent modes, $h_{b}^{\ae},\; h_{X}^{\ae}$ and $h_{Y}^{\ae}$. Comparing to $h_+^{\ae}$ and $h_\times^{\ae}$, these extra modes are suppressed, respectively, by a factor, { $c_{14} \lesssim 10^{-5}$} and $c_{13} \lesssim 10^{-15}$ {\cite{Lin}}. The longitudinal mode $h_L^{\ae}$ is proportional to the breathing mode $h_b^{\ae}$. 

In the following, let us first consider the case,
\bqn
\lb{4.1c}
c_1&=& 4\times 10^{-5}, \;\; c_2=9\times 10^{-5}, \nb\\
 c_3 &=& -c_1, \;\; c_4=-2\times 10^{-5},
\eqn
which satisfy all the theoretical and observational constraints given by Eq.(\ref{2.8b}) (See also \cite{OMW18}). Note that for this choice we have $c_{13} = 0$, and then the two modes $h_{X}^{\ae}$ and  $h_{Y}^{\ae}$ vanish identically, 
\bq
\lb{4.1d}
h_{X}^{\ae} = h_{Y}^{\ae} = 0, \; (c_{13} = 0).
\eq
So, in the rest of this paper, we shall not consider them any further. 

In Figs. \ref{fig2} and \ref{fig3}, we plot the two polarization modes $h_+^{\ae}$ and $h_\times^{\ae}$, while in Figs. \ref{fig4} and \ref{fig5}, we plot their DFTs. From these figures, it can be seen clearly that these two modes are almost identical to the ones given in GR, after all the constraints of $\ae$-theory are taken into account. In fact, we have
\bq
\lb{4.1da}
\frac{h_{+, \times}^{\ae}}{h_{+, \times}^{GR}} = 1 - \frac{1}{2} c_{14}.
\eq
The differences between  {\ae}-theory and GR are determined by $c_{14}$, 
\bq
\lb{4.1db}
\Delta h^{\ae}_{+, \times} \equiv \frac{h^{\ae}_{+, \times} - h^{GR}_{+, \times}}{h^{GR}_{+, \times}}=- \frac{1}{2} c_{14}.
\eq
Recall that $c_{14} \lesssim 2.5 \times 10^{-5}$. Therefore, the signals of these two modes in ${\ae}$-theory and GR are overlapping, their frequencies  are  precisely the same, as shown in Figs. \ref{fig4} and \ref{fig5}. Similarly, the differences in frequency domain can be defined as
 \bq
 \lb{4.1dc}
 \Delta |\tilde{h}^{\ae}_{+, \times}| \equiv \left( |\tilde{h}^{\ae}_{+, \times}| - |\tilde{h}^{GR}_{+, \times}|\right) / |\tilde{h}^{GR}_{+, \times}|.
\eq

\begin{figure}[h!]
\includegraphics[width=\linewidth]{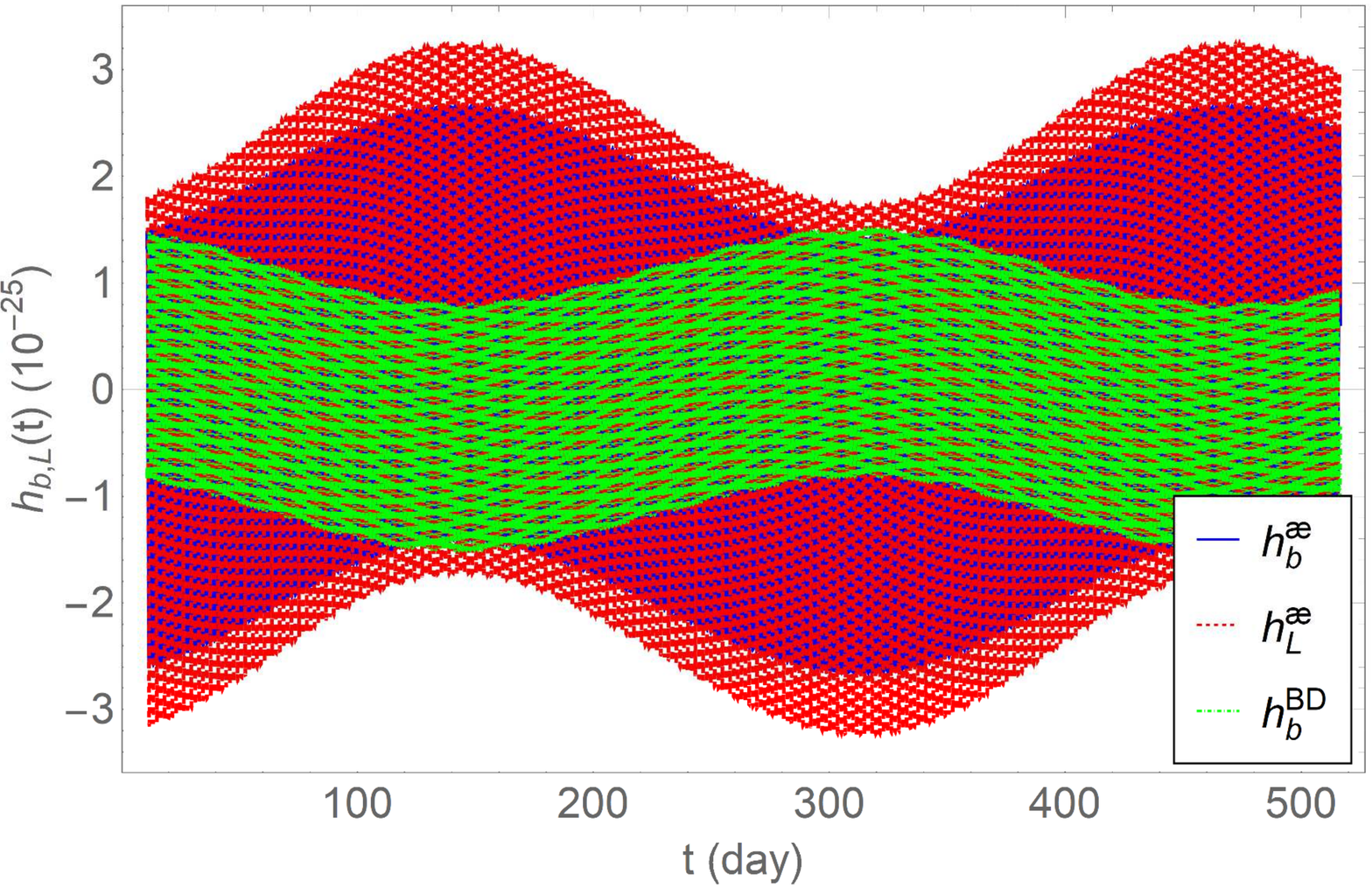} \\
\includegraphics[width=\linewidth]{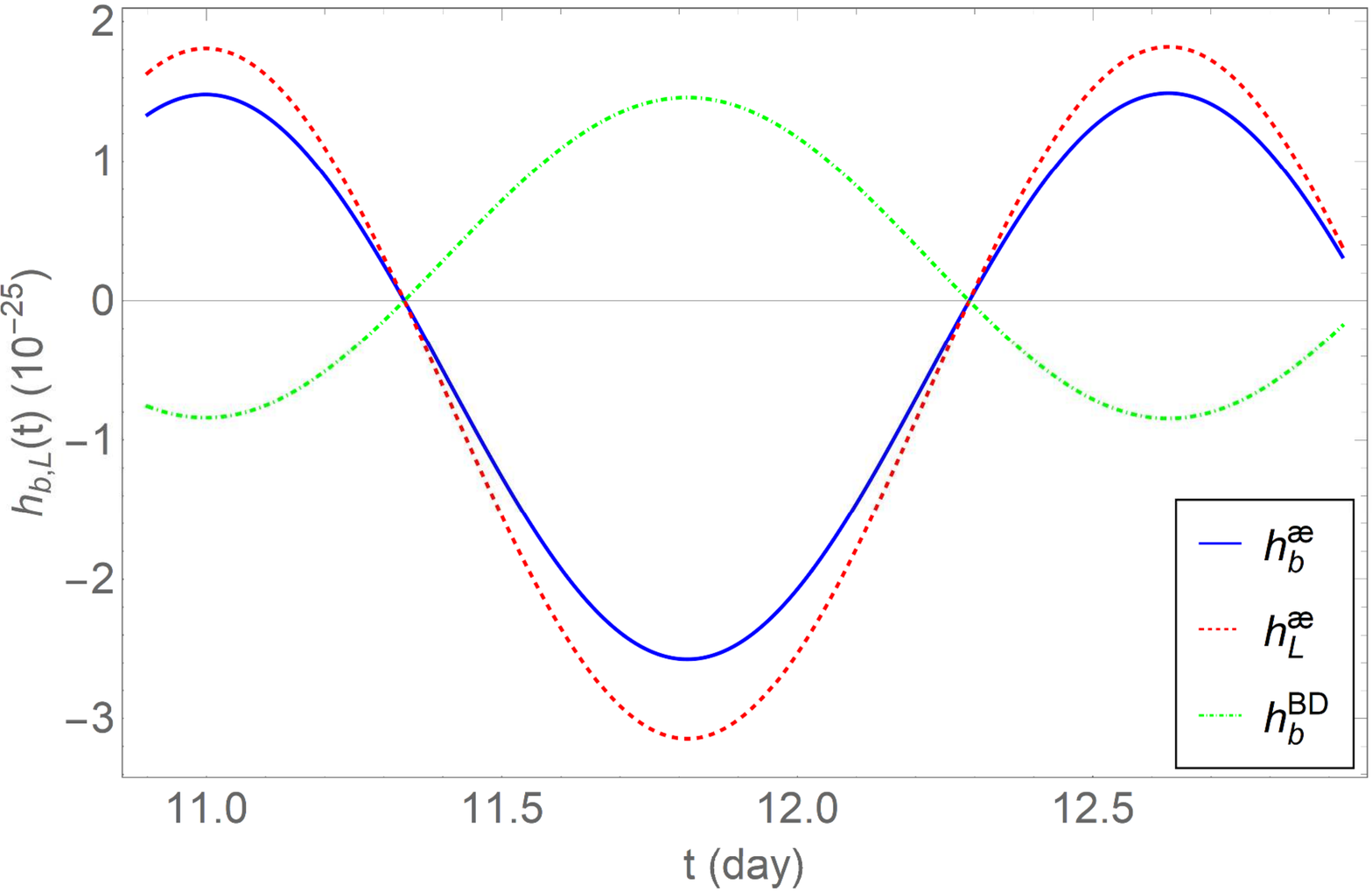} 
\caption{The breathing ($h_{b}$) and longitudinal ($h_{L}$) polarizations in {\ae}-theory and the breathing polarization in BD gravity for about 500 days with another plot for about 2 days.} 
\label{fig6}
\end{figure}

In Fig. \ref{fig6}, we plot $h_{b}^{\ae}$ and $h_{L}^{\ae}$, which are about three orders lower than $h_+^{\ae}$ and $h_\times^{\ae}$. In Fig. \ref{fig7}, we plot the DFT of the $h_b$ and $h_L$. It is remarkable that now the two peaked  frequencies are approxiamtely equal to the outer and inner orbital frequencies. 
 \bqn
 \lb{4.1e}
 f^{b, L}_1 &=& 0.045772 \mu \mbox{Hz} \simeq  1.3 f_o,\nb\\
 f^{b, L}_2 &=& 7.0947 \mu {\mbox{Hz}} \simeq 1.0 f_i, 
 \eqn
 where $f_o$ and $f_i$ are given by Eq.(\ref{3.5c}). $f^{b, L}_1$ and $f^{b, L}_2$ are almost equal to the outer and inner orbital frequencies of the triple system.
 
Recently,  we find that, for binary systems in $\ae$-theory, the polarization modes $h_b$ and $h_L$ all contain two frequencies, one is equal to the binary's orbital frequency and the other is twice the  orbital frequency  \cite{Zhao19}. This confirms the above result since J0337 can be considered a hierarchical system consisting of two binaries. Again, the reason that  $f^{b, L}_1$ and $f^{b, L}_2$ are not exactly equal to the outer and inner orbital frequencies of the triple system is due to the influence of the third component of the triple system, rather than the predictions of the $\ae$-theory itself.

\begin{figure}[h!]
\includegraphics[width=\linewidth]{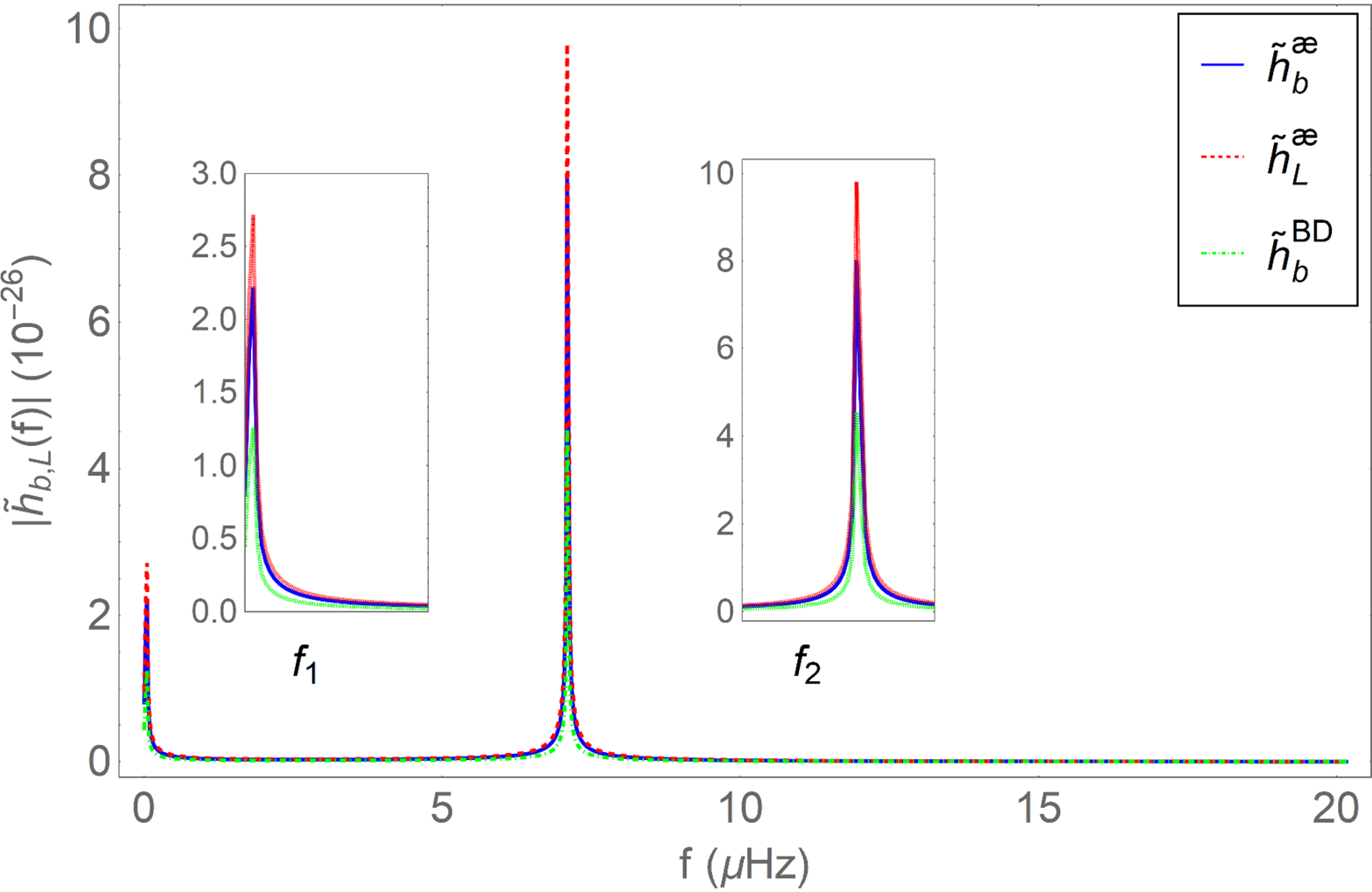} 
\caption{DFT of the breathing and longitudinal polarizations in {\ae}-theory and the breathing polarization in BD gravity, where two peaked frequencies have been marked. The inserted images show the tiny differences at two peaked frequencies among theories.  $f_1= 0.045772\mu$Hz, $f_2=7.0947\mu$Hz.}
\lb{fig7}
\end{figure}

To see further the dependence of $h_{b, L}^{\ae}$ on $c_i$'s, let us first note that  
\bqn
\lb{B.1}
h_b^{\ae} &\simeq& \frac{c_S}{1 + c_S} h_L^{\ae} \nb\\
& \simeq& \frac{G_{\ae}}{2R} \left\{3c_{14} \ddot{Q}_{ij}e_Z^i e_Z^j 
 - \sqrt{\frac{c_{14}}{c_2}}\Sigma_i e^i_Z + 2c_{14}\ddot{I}\right\}, ~~~~~~~
\eqn
where $c_{14}/c_2 \lesssim 1$, as it can be seen from Eq.(\ref{2.8b}).

From above equation, we see that $h_{b, L}^{\ae}$ are dependent on $c_2$ and $c_{14}$ mainly. To see the effects explicitly,  in the following we consider two more cases.

\begin{figure}[h!]
\includegraphics[width=\linewidth]{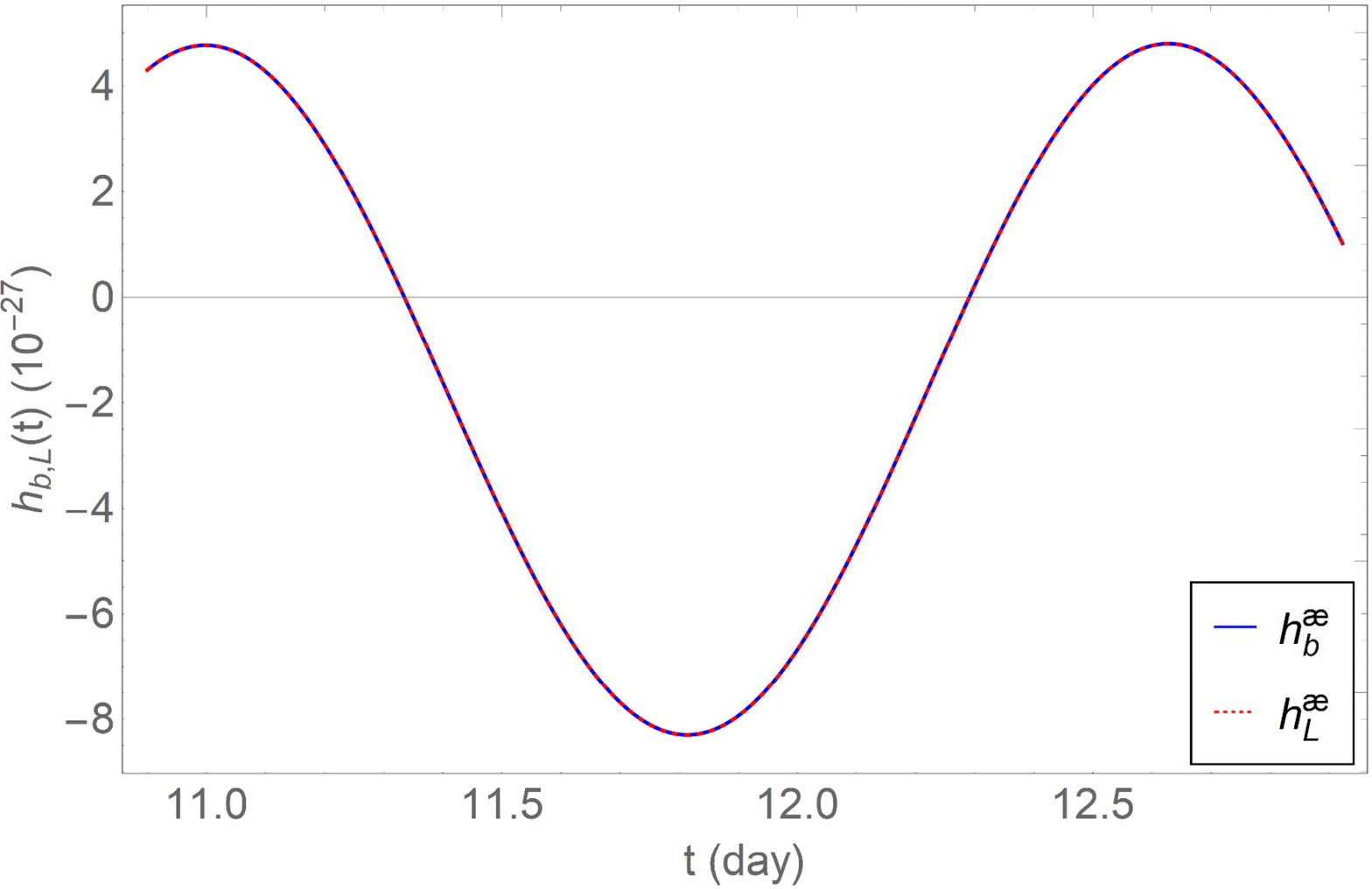} 
\caption{The breathing and longitudinal polarizations in ${\ae}$-theory for different choice of parameters $c_i$'s given by Eq.(\ref{B1}).}
\label{fig8}
\end{figure}

\begin{figure}[h!]
\includegraphics[width=\linewidth]{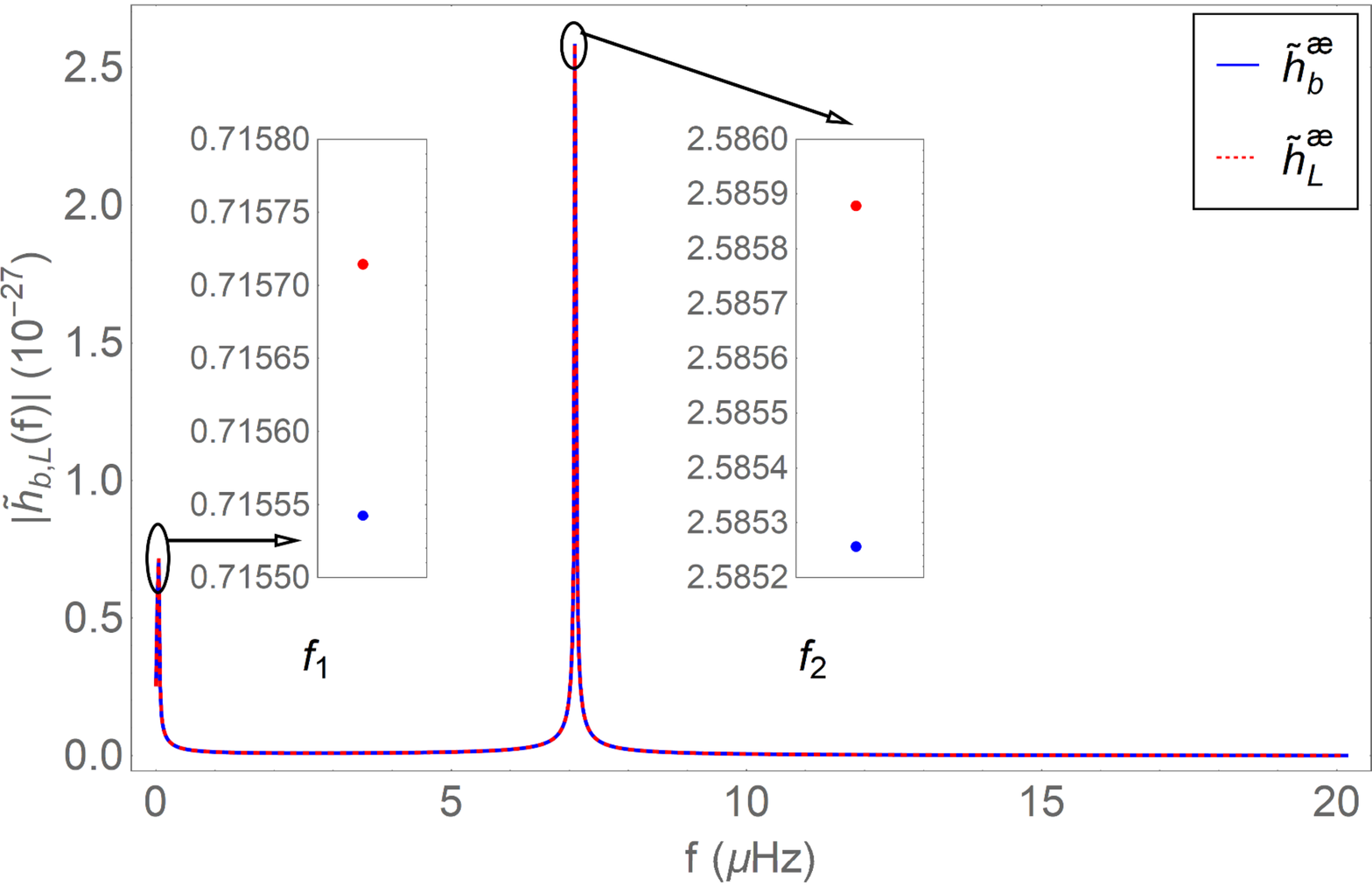} 
\caption{DFT of  the breathing and longitudinal polarizations in {\ae}-theory for different choice of parameters $c_i$'s given by Eq.(\ref{B1}). $f_1= 0.045772\mu$Hz, $f_2=7.0947\mu$Hz.}
\label{fig9}
\end{figure}

In the first one,  $c_2$ is chosen to be different from Eq. (\ref{4.1c}), now we have
\bqn
\lb{B1}
c_1&=& 4\times 10^{-5}, \;\; c_2=0.095, \nb\\
 c_3 &=& -c_1, \;\; c_4=-2\times 10^{-5}.
\eqn
In Figs. \ref{fig8} and \ref{fig9} we plot out the breathing ($h_{b}^{\ae}$) and longitudinal ($h_{L}^{\ae}$) polarization modes for this case. Comparing them, respectively,  with  Figs. \ref{fig6} and \ref{fig7}, we find that the 
amplitudes of $h_b^{\ae}$ and $h_L^{\ae}$ decrease  when $c_2$ increases, whereas their frequencies  remain the same.

\begin{figure}[h!]
\includegraphics[width=\linewidth]{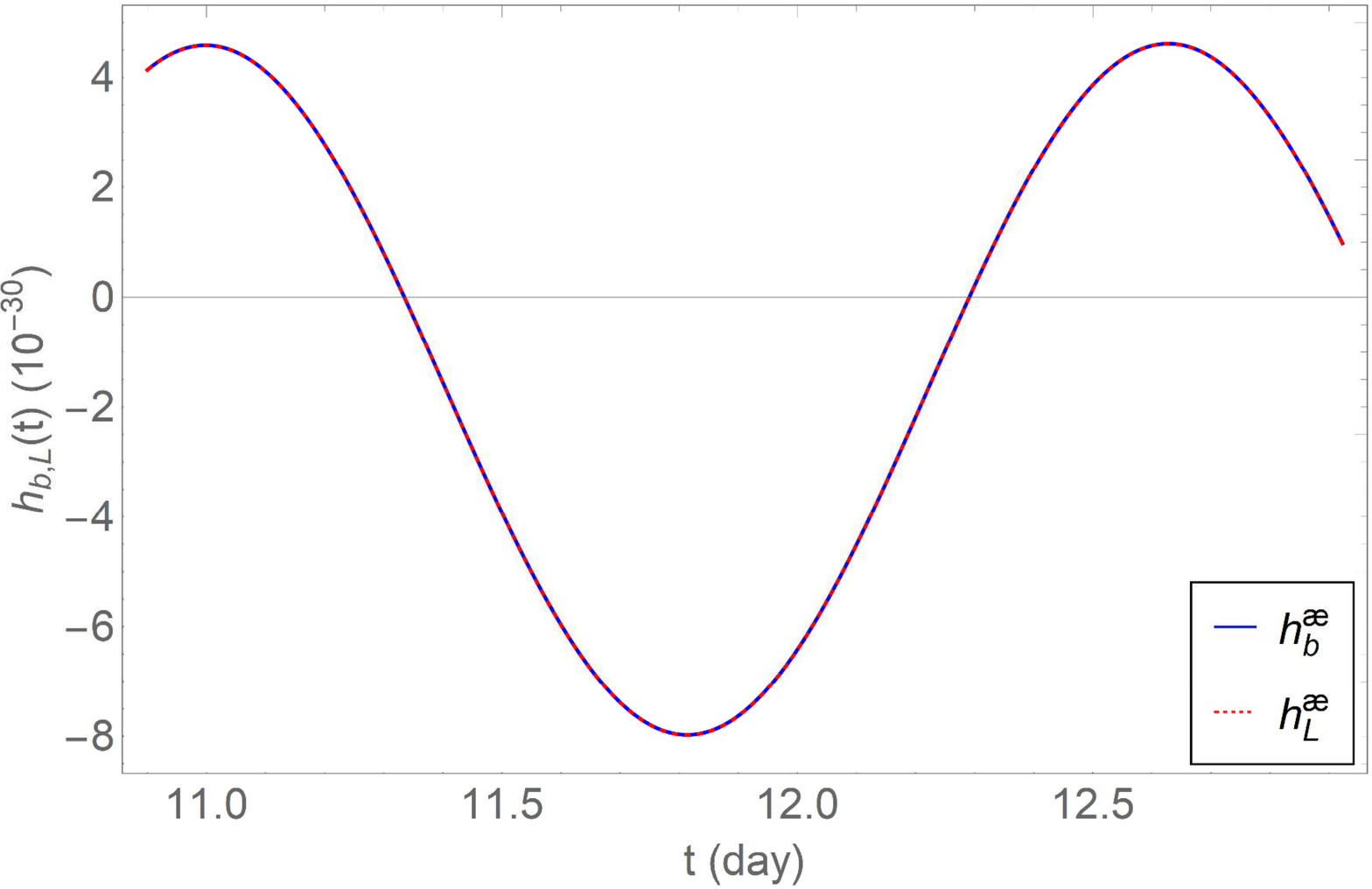} 
\caption{The breathing and longitudinal polarizations in {\ae}-theory for different choice of parameters $c_i$'s given by Eq.(\ref{B2}).}
\label{fig10}
\end{figure}

\begin{figure}[h!]
\includegraphics[width=\linewidth]{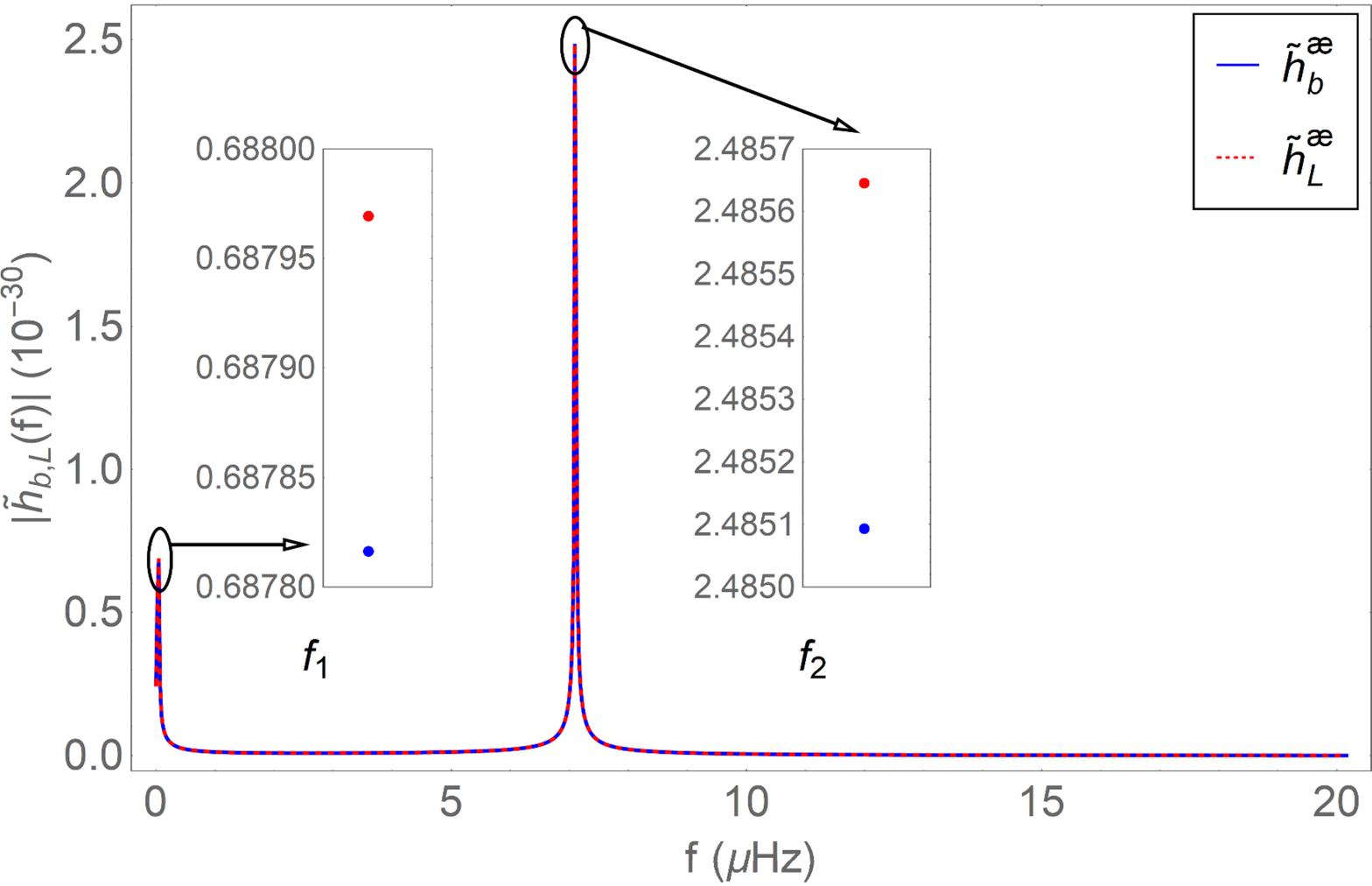} 
\caption{DFT of  the breathing and longitudinal polarizations in {\ae}-theory for different choice of parameters $c_i$'s given by Eq.(\ref{B2}). $f_1= 0.045772\mu$Hz, $f_2=7.0947\mu$Hz.}
\label{fig11}
\end{figure}

On the other hand,  $c_{14}$ is chosen to be different from Eq. (\ref{4.1c}), the parameters are
\bqn
\lb{B2}
c_1&=& 4\times 10^{-8}, \;\; c_2=9\times10^{-5}, \nb\\
 c_3 &=& -c_1, \;\; c_4=-2\times 10^{-8},
\eqn
we plot out the corresponding $h_{b}^{\ae}$ and $h_{L}^{\ae}$ modes in Figs. \ref{fig10} and \ref{fig11}. Comparing them with Figs. \ref{fig6} and Fig. \ref{fig7}, we find that the amplitudes
 of $h_b^{\ae}$ and $h_L^{\ae}$ decrease when $c_{14}$ decreases, whereas their frequencies  stay the same, as it is expected from Eq.(\ref{B.1}).

\subsection{Gravitational Waveforms and Their Polarizations in BD Gravity}

The metric perturbation and scalar field perturbation are given by \cite{Will},
\bqn
\lb{3.9}
\theta^{ij} &=& \frac{2}{\phi_0} \frac{1}{R} \frac{d^2}{dt^2} \sum\limits_{a=1}^3 m_a x^i_a x^j_a, \nb\\
\varphi^{BD}     &=& \frac{4}{R} (N_i \dot{M}_1^i + \frac{1}{2} N_i N_j \ddot{M}_2^{ij}),
\eqn
where $\phi_0$ is the value of the {BD} {scalar} field in the Minkowski background, which satisfies $\phi_0 =  (4+2 \omega_{BD})/[(3+2 \omega_{BD})G_N]$ \cite{Zhang17},  and
\bqn
\lb{2.7}
M_1^i    &= \frac{1}{6 + 4 \omega_{BD}} \sum\limits_{a=1}^3 m_a (1 - 2 s_a) x^i_a, \nb\\
M_2^{ij} &= \frac{1}{6 + 4 \omega_{BD}} \sum\limits_{a=1}^3 m_a (1 - 2 s_a) x^i_a x^j_a,
\eqn
where $\omega_{BD}$ is the BD parameter of the theory. In this paper, we choose sensitivities such that $s_1$ (for pulsar) $= 0.2$, $s_2$ (for inner WD) $= 0$, $s_3$ (for outer WD) $= 0$ and the coupling constant {$\omega_{BD} = 10^5$ \cite{Archibald18}}.
Note that in writing the above expressions, we had dropped the non-propagating terms in $\varphi^{BD}$. Then,  the components $R_{0i0j}$ of the Riemann tensor can be cast in the form, 
\bq
\lb{3.10}
R_{0i0j} = - \frac{1}{2} \frac{d^2}{dt^2} \left[ \theta_{ij}^{TT} - \frac{\varphi^{BD}}{\phi_0} \left( \delta_{ij} - N_i N_j \right) \right].
\eq 	
Then, it can be shown that there are only three independent polarization modes, given, respectively, by  
\bqn
\lb{3.11}
h_+^{BD}      &=& \frac{1}{2}e_{+}^{ij}   \theta_{ij}, \;\;
h_\times^{BD} = \frac{1}{2}e_{\times}^{ij} \theta_{ij}, \nb\\
h_b^{BD}      &=& -\frac{\varphi^{BD}}{\phi_0},
\eqn
which are plotted out, respectively, in Figs. \ref{fig2}, \ref{fig3} and \ref{fig6}, while their DFTs are plotted out in  Figs. \ref{fig4}, \ref{fig5} and \ref{fig7}.  
From these figures, it can be seen that  the two polarization modes $h_+^{BD}$ and $h_\times^{BD}$ are overlapping with those in GR and 
$\ae$-theory, due to the observational constraints on the $\omega_{BD}$.
In fact, we have
\bq
\lb{3.11a}
\frac{h_{+, \times}^{BD}}{h_{+, \times}^{GR}} = \frac{3+2\omega_{BD}}{4+2\omega_{BD}}.
\eq

The differences between  BD gravity and GR are determined by $\omega_{BD}$, 
\bq
\lb{3.11b}
\Delta h^{BD}_{+, \times} \equiv \frac{h^{BD}_{+, \times} - h^{GR}_{+, \times}}{h^{GR}_{+, \times}}=\frac{3+2\omega_{BD}}{4+2\omega_{BD}} -1.
\eq
Recall that $\omega_{BD} \sim {\cal O}(10^5)$. Therefore, the signals of these two modes in BD gravity and GR are overlapping, their frequencies  are  precisely the same, as shown in Figs. \ref{fig4} and \ref{fig5}. Similarly the differences in frequency domain can be defined as
\bq
\lb{3.11c}
\Delta |\tilde{h}^{BD}_{+, \times}| \equiv \left( |\tilde{h}^{BD}_{+, \times}| - |\tilde{h}^{GR}_{+, \times}|\right) / |\tilde{h}^{GR}_{+, \times}|.
\eq 
From Fig. \ref{fig6}, the breathing mode ($h_b^{BD}$) is different from $\ae$-theory. From Fig. \ref{fig7}, its DFT has also two peaked frequencies and are equal to those of $\ae$-theory, i.e., breathing mode in BD gravity only has first harmonics of orbital phase. In contrast to the polarization modes $h_{+}$ and $h_{\times}$, which have second harmonics of orbital phase.

\section{Radiation Power}
\renewcommand{\theequation}{3.\arabic{equation}}
\setcounter{equation}{0}
 	
In {GR},  the total radiation  power is given by \cite{MM, Dmitra},
\bq
\lb{2.1}
P^{GR} = \frac{G_N}{5} \left\langle \dddot{Q}_{ij} \dddot{Q}_{ij} \right\rangle,
\eq
where   $Q_{ij}$ is the mass quadrupole moment defined in Eq.(\ref{4.1b})  and the angular brackets denote the time average. \footnote{However, since we are considering periodic GWs, we will not take this time average in the relevant  plots. Otherwise, it will  be zero for such periodic waves. The same will also apply to the cases of $\ae$-theory and BD gravity.}. 
Note that in this section we   shall not  distinguish the time $t$ and its corresponding retarded time. Strictly speaking, all the quantities should be evaluated at the retarded time. However, it is not necessary    for our current purpose.
 
\begin{figure}[h!]
\includegraphics[width=\linewidth]{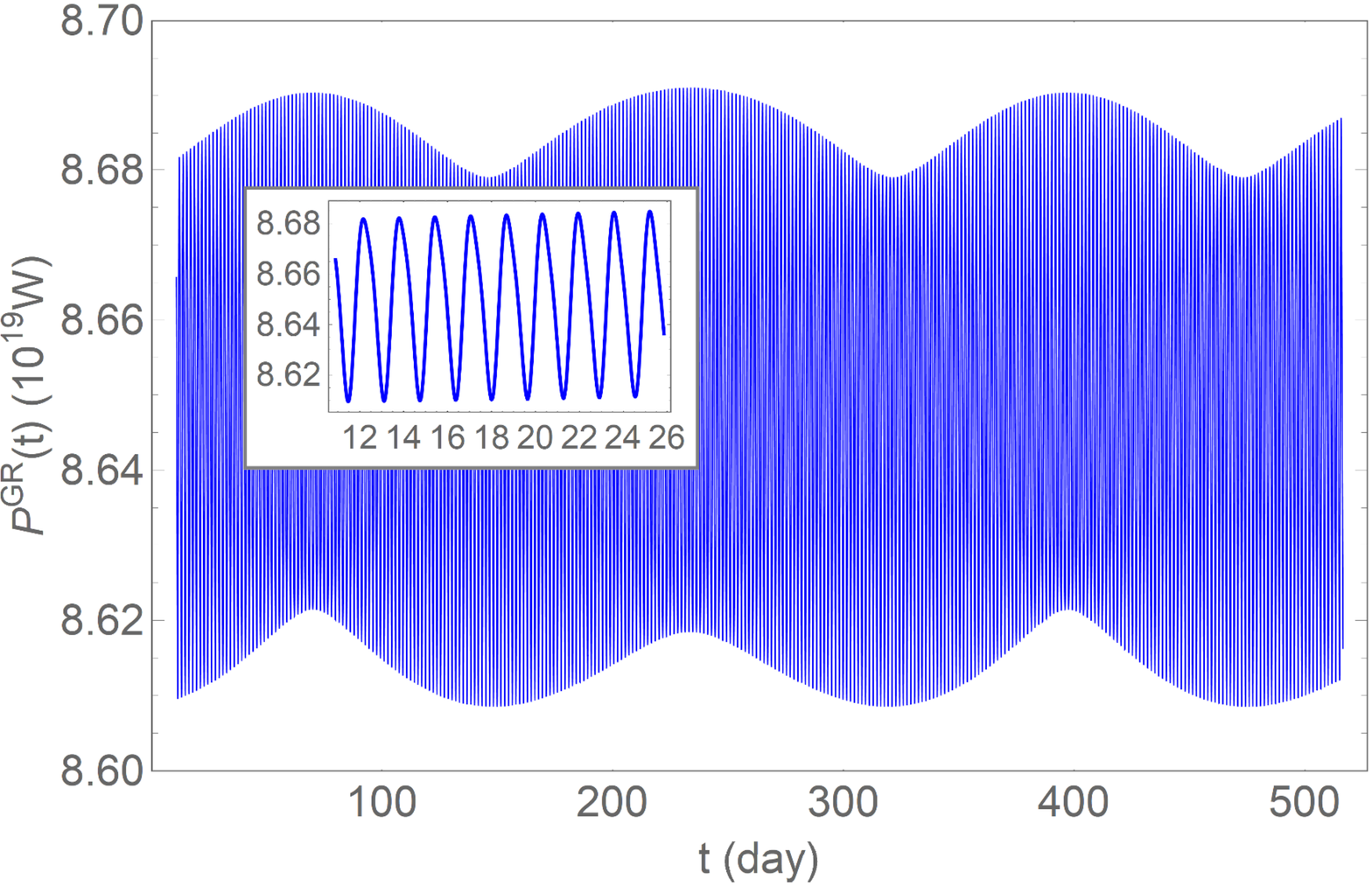} 
\caption{The instantaneous (time un-averaged) radiation power in GR for about 500 days with an inserted  plot  for about 14 days.}
\label{fig12}
\end{figure}

The reference frame is chosen such that the inclination is $39.25^\circ$, where the inclination is the angle of the orbital plane relative to the ($x, y$)-plane   perpendicular to the line-of-sight from Earth to the pulsar.
In Fig. \ref{fig12}, we plot the radiation  power in GR for about 500 days, where the inserted image shows the details from day 12 to day 26.   

In $\ae$-theory, from \cite{Foster,Lin} we find that
\bq
\lb{2.3}
P^{\ae} = G_N\left\langle \frac{\cal A}{5}\dddot{Q}_{ij}\dddot{Q}_{ij}+{\cal B}\dddot{I}\dddot{I}+{\cal C}\dot{\Sigma}_i\dot{\Sigma}_i\right\rangle,
\eq
where $\Sigma_i$ is defined as
\bqn
\lb{2.4}
\Sigma_i &= \left(\alpha_1-\frac{2}{3}\alpha_2\right)\sum\limits_{a}\left(v^i_a \Omega_a\right),
\eqn
and 
\bqn
\lb{2.5}
{\cal A}&\equiv& \left(1-\frac{c_{14}}{2}\right)\left(\frac{1}{c_T}+\frac{2c_{14}c_{13}^2}{(2c_1-c_{13} c_-)^2c_V}\right. \nb\\
        && \left.~~~~~~~~~~~~~~~ +\frac{3(Z-1)^2 c_{14}}{2(2-c_{14})c_S}\right), \nb\\
{\cal B}&=& \frac{Z^2 c_{14}}{8c_S}, \quad \nb\\
{\cal C}&\equiv& \frac{2}{3 c_{14}}\left(\frac{2-c_{14}}{c_V^3}+\frac{1}{c_S^3}\right), \nb\\
 Z &\equiv& \frac{(\alpha_1 - 2\alpha_2)(1-c_{13})}{3(2c_{13} - c_{14})},
\eqn
with  \cite{Foster06},
\bqn
\lb{2.3aa}
\alpha_1 &=& -  \frac{8\left(c_1c_{14} - c_{-}c_{13}\right)}{2c_1 - c_{-}c_{13}}, \nb\\
\alpha_2 &=&   \frac{1}{2}\alpha_1  + \frac{\left(c_{14}- 2c_{13}\right)\left(3c_2+c_{13}+c_{14}\right)}{c_{123}(2-c_{14})}. 
\eqn
 Here  $v^i_a\equiv \dot{x}^i_a$ is the velocity of $a$-th body along $x^i$-direction, $\Omega_a$ is the binding energy of $a$-th body.
For J0337 \cite{Ransom14}, we have  $\Omega_1$ (for pulsar) =$ -2.56955 \times 10^{46}$ J, $\Omega_2$ (for inner WD)=$-9.75554 \times 10^{40}$ J, $\Omega_3$ (for outer WD)=$-2.12650 \times 10^{42}$ J.
where $c_T$, $c_V$ and $c_S$ are the speeds of the tensor, vector and scalar modes, given by Eq.(\ref{CSs}) in Appendix A, in which the definitions  $c_{ijk} \equiv c_i +c_j +c_k$ and $c_{ij} \equiv c_i+c_j$ are given. 

\begin{figure}[h!]
\includegraphics[width=\linewidth]{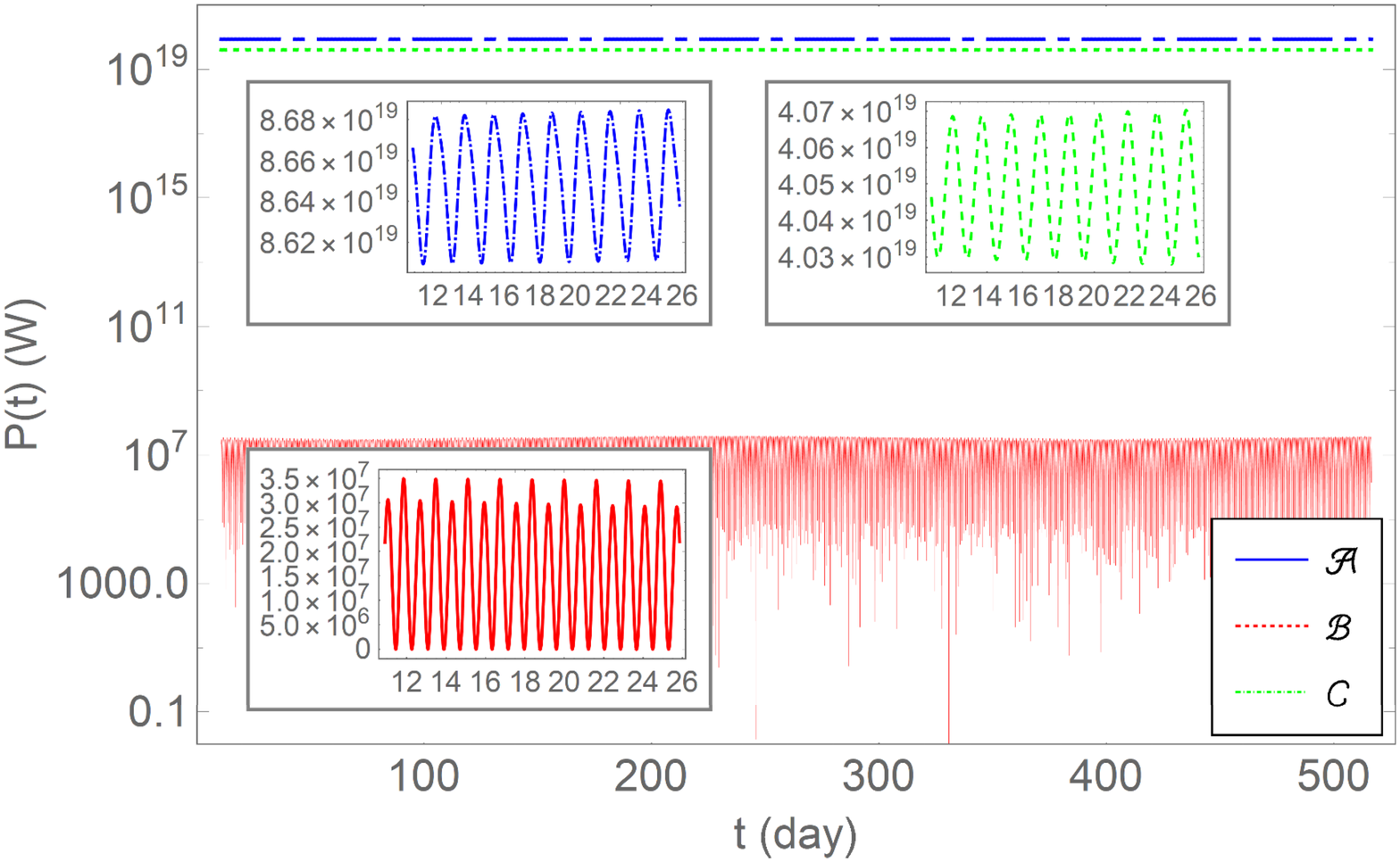} 
\caption{The instantaneous (time un-averaged) radiation power in $\ae$-theory. Here the ${\cal A}$, ${\cal B}$ and ${\cal C}$ parts are plotted separately. The inserted  plots shows only 14 days.}
\label{fig13}
\end{figure}

In Fig. \ref{fig13}, we plot the radiation power in ${\ae}$-theory of the parts ${\cal A}$, ${\cal B}$ and ${\cal C}$ separately,   for about 500 days. Again, the inserted  images  are from day 12 to day 26. Note that at every moment during the 252 days in the plot, the ${\cal A}$ part of ${\ae}$-theory is quite close to that of GR with the {relative} difference proportional to $c_{14}$ \cite{Lin},
\bq
\lb{2.3ab}
\frac{P_{\cal A}^{\ae}}{P^{GR}} -1\simeq  {\cal O}(c_{14}) \lesssim {\cal O}(10^{-5}).
\eq

From this figure, it is also clear that the dipole part ${\cal C}$ has almost the same amplitude as that of the quadrupole part ${\cal A}$, while the monopole part ${\cal B}$ is suppressed by a factor $c_{14}$ \cite{Lin}. 
The large magnitude of the dipole contribution  $\cal{C}$   seemingly contradicts to the analysis given in  \cite{Lin}. In particular, Eq.(3.13) in  \cite{Lin}
shows that ${\cal{W}}_{\cal{C}}^{\mbox{NS}}/{\cal{W}}_{\cal{A}}^{\mbox{NS}} \simeq 10^{-2}$, where
\bqn
\lb{2.3ac}
{\cal{W}}_{\cal{A}} &\equiv&  \frac{8}{15}{\cal{A}}\left(12v^2 - 11\dot{r}^2\right), \nb\\
{\cal{W}}_{\cal{C}} &\equiv& {\cal{C}}\Sigma^2,
\eqn
where ${\cal{A}},\; {\cal{C}}$ and $\Sigma$ are all given explicitly in  \cite{Lin}. However, in deriving   ${\cal{W}}_{\cal{C}}^{\mbox{NS}}/{\cal{W}}_{\cal{A}}^{\mbox{NS}} \simeq 10^{-2}$ we assumed that $ {\cal{O}}\left(v^2\right) \simeq 10^{-5}$, while in the case of J0337 we find that the relative
velocities of the inner binary system are of the order of  $ {\cal{O}}\left(v^2\right) \simeq 10^{-7}$. After this is taken into account, we find that  ${\cal{W}}_{\cal{C}}/{\cal{W}}_{\cal{A}}\simeq  {\cal{O}}(1)$
for the current triple system. 

 It is remarkable to note that,  with the multi-band gravitational wave astronomy \cite{AS16}, joint observations of GW150914-like by LIGO/Virgo/KAGRA and LISA will improve bounds on dipole 
emission from black hole binaries by six orders of magnitude relative to current constraints \cite{BYY18}. Thus, it is very promising that the third generation of detectors, both space-borne and ground-based, 
could provide severe constraints on   $\ae$-theory.

In BD gravity, following \cite{Will} we obtain 
\bqn
\lb{2.6}
P^{BD} &=& P_1^{BD}+P_2^{BD},
\eqn
where
\bqn
\lb{2.6a}
P_1^{BD} &=& \frac{1}{\phi_0} \left\langle \frac{1}{5} \dddot{Q}_{ij} \dddot{Q}_{ij}   \right\rangle,\nb\\
P_2^{BD} &=& \frac{1}{\phi_0} \left\langle \frac{2 \omega + 3}{\pi} \left[\frac{4 \pi}{3} \ddot{M}_1^i \ddot{M}_1^i \right. \right. \nb\\
       && + \left. \left. \frac{\pi}{12} \left( \dddot{M}_2^{ii} \dddot{M}_2^{jj} + 2 \dddot{M}_2^{ij} \dddot{M}_2^{ij} \right) \right] \right\rangle,
\eqn
where ${M}_1^i$ and ${M}_2^{ij}$ are defined by Eq.(\ref{2.7}). Note that in writing down the above expressions,  we had dropped the non-propagating terms.  

\begin{figure}[h!]
\includegraphics[width=\linewidth]{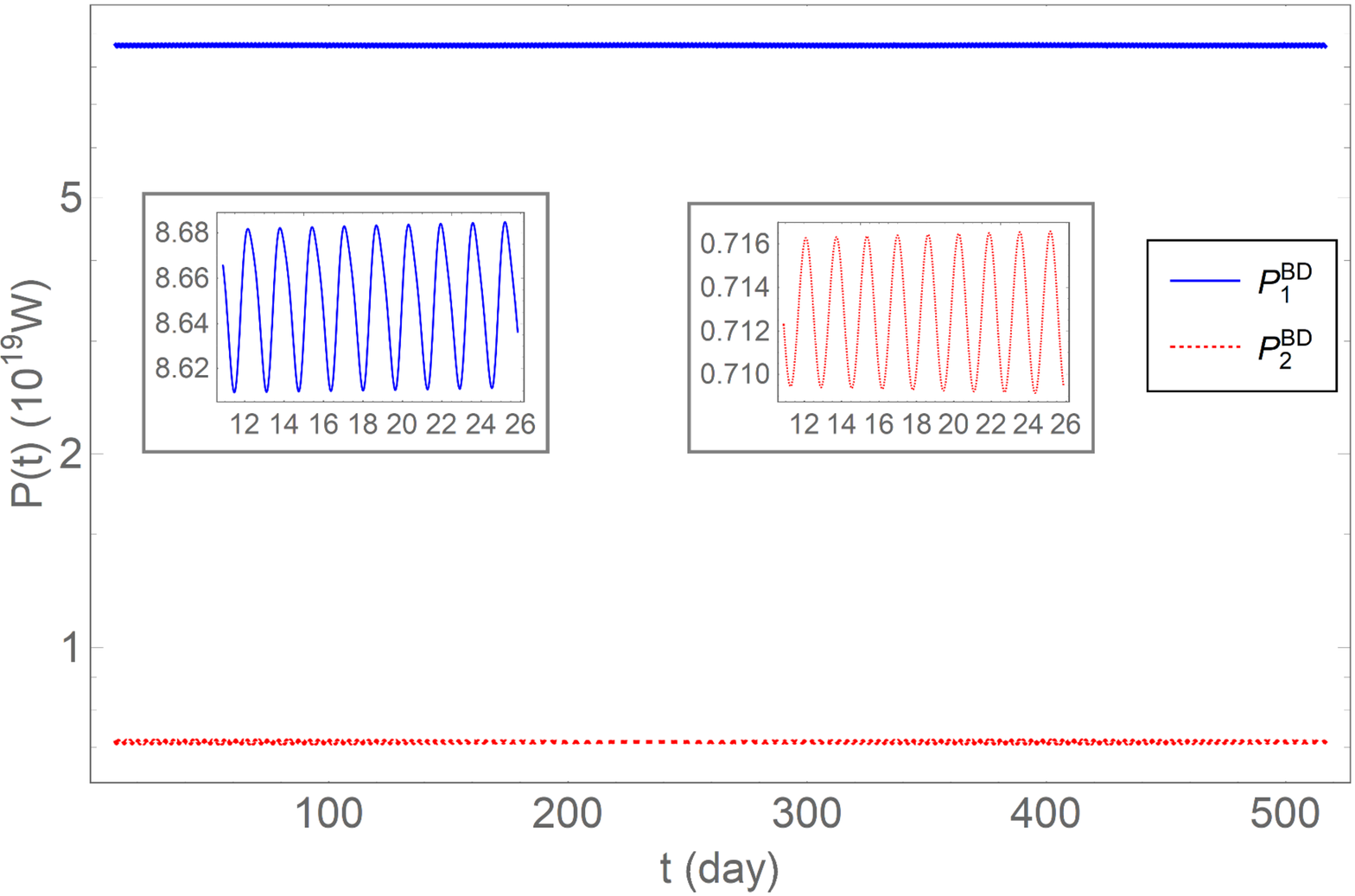} 
\caption{The instantaneous (time un-averaged) radiation power in BD gravity. Here the part 1 and 2 are plotted,  separately. The inserted  plots shows only 14 days}
\label{fig14}
\end{figure}

In Fig. \ref{fig14}, we plot the radiation power in BD gravity for about 500 days, where the inserted image  shows the details only from day 12 to day 26. Note that at every moment during the 500 days, the first part of BD is quite close to that given in GR. In fact, we find that
\bq
\lb{2.6b}
 {\frac{P_1^{BD}}{P^{GR}} -1 \simeq {\cal O}\left(\omega_{BD}^{-1}\right) \lesssim {\cal O}(10^{-5}).}
 \eq
 
\section{Conclusions}
\renewcommand{\theequation}{4.\arabic{equation}}
\setcounter{equation}{0}

In this  paper, we studied the gravitational waveforms, their polarizations and Fourier transforms, as well as the radiation powers of the relativistic triple systems PSR J0337 + 1715, observed in 2014 \cite{Ransom14}.
This system   consists of an inner  binary and a third companion.  The inner binary consists 
 of a pulsar  with mass  $m_1 = 1.44 M_{\bigodot}$ and a white dwarf with mass  $m_2 = 0.20 M_{\bigodot}$ in a 1.6 day orbit.  The outer  binary consists of the inner binary and a second dwarf with mass    
 $m_3 = 0.41 M_{\bigodot}$ in a  327 day orbit.  The two orbits are very circular with its eccentricities $e_I \simeq 6.9\times 10^{-4}$ for the inner binary and $e_O \simeq 3.5  \times 10^{-2}$ for the outer orbit.  
 The two orbital planes  are remarkably  coplanar with an inclination  $\lesssim 0.01^{o}$ [See Fig. \ref{fig1}].

Our studies were carried out in three different theories, GR, BD gravity, and $\ae$-theory. In GR, only the tensor mode exists, so a GW has only two polarization modes, the so-called, plus ($h_{+}$) and cross ($h_{\times}$) modes. 
Their amplitudes and Fourier transforms are plotted in Figs. \ref{fig2} - \ref{fig4}, from which it can be seen that their amplitude is about $10^{-23}$, while their frequencies are peaked in two locations, 
$ f^{+, \times} _1 = 0.068658 \mu$Hz {(for the outer orbit)} and $ f^{+, \times}_2 = 14.212 \mu$Hz {(for the inner orbit)}, respectively. These  are  about two times of the inner and outer   
orbital frequencies of the triple system, and agree well with the GR predictions \cite{MM}. 

In $\ae$-theory, all the six polarization modes are different from zero, but the beating ($h_b$) and longitudinal ($h_L$) modes are not independent and are related to each other by Eq.(\ref{3.8}). In comparing with $h_{+}$ and $h_{\times}$, however, 
they are suppressed by a factor $c_{14}$ {which is} observationally restricted to $c_{14} \lesssim 10^{-5}$ \cite{OMW18}.  A somehow surprising result is that the two peaked frequencies are not two times of those of the inner and outer   
orbital frequencies of the triple system. Instead, they are about equal to them, $ f^{b, L}_{1} = 0.045772 \mu$Hz (for the outer orbit) and $f^{b, L}_{2} = 7.0947 \mu$Hz (for the inner orbit) [Cf. Fig. \ref{fig7}]. 

The effects of the parameters $c_i$'s on $h_b^{\ae}$ and $h_L^{\ae}$ were also studied in detial, and in particular we found that their amplitudes are weekly dependent on the choices of these parameters, while the frequencies of their Fourier transforms remain the same. 

The other two independent polarization modes in $\ae$-theory are the vector modes, $h_{X}$ and $h_{Y}$, which are all proportional to $c_{13}$.
The current observations from GW170817   \cite{GW170817}  and  GRB 170817A \cite{GRB170817}  on   the speed of the tensor mode requires $c_{13} \lesssim 10^{-15}$. Therefore, these two modes are highly restricted by the limit
of the speed of the tensor mode.     

We also studied the radiation power due to the tensor, vector and scalar modes in $\ae$-theory, and three different parts were plotted in Fig. \ref{fig9}. The amplitude of the quadrupole part (${\cal{A}}$), contributed from all of these 
three modes, tensor, vector and scalar  \cite{Foster,Lin} is quite comparable with that of GR. But, the monopole (${\cal{B}}$) part has contributions only from the scalar mode, while the dipole (${\cal{C}}$) part has contributions from both the
scalar and vector modes, but does not have any contributions from the tensor mode, as expected \cite{Foster,Lin}. The monopole part is suppressed by a factor $c_{14} \lesssim {\cal{O}}\left(10^{-5}\right)$, 
but the dipole part is almost in the same order of the quadrupole part. With the arrival of the multi-band gravitational wave astronomy \cite{AS16}, joint observations of GW150914-like by LIGO/Virgo/KAGRA and LISA will improve serve constraints on  the dipole 
emission  \cite{BYY18}. Thus,  the multi-band gravitational wave astronomy will provide a very   promising direction to constrain $\ae$-theory.

We also carried out similar studies in BD gravity, and the relevant quantities were plotted in Figs. \ref{fig2} - \ref{fig7} and \ref{fig10}. Due to the severe observational constraint on the BD parameter $\omega_{BD} \gtrsim 10^{5}$, we did not find significant deviations from GR, except that the frequency of the breathing mode $h_b$ is also peaked in two locations, $ f^{b}_{1} = 0.045772 \mu$Hz and $f^{b}_{2}= 7.0947 \mu$Hz, which are about the outer and inner orbital frequencies of the triple system, quite similar to that in  $\ae$-theory  [Cf. Fig. \ref{fig7}] but with a low amplitude.

\section*{Acknowledgments}
We would like to thank Dr. Lijing Shao for providing us the original data of PSR J0337+1715.
This work is supported in part by the National Natural Foundation of China (NNSFC) with the grant numbers:  Nos. 11603020, 11633001, 11173021, 11322324, 11653002, 11421303, 11375153, 11675145, 
11675143,  11105120, 11805166, 11835009, 11690022, 11375247, 11435006, 11575109,  11647601, and No. 11773028.

\section*{Appendix A: Einstein-aether Theory}
\renewcommand{\theequation}{A.\arabic{equation}}
\setcounter{equation}{0}

In this appendix, we shall give a brief introduction to Einstein-aether Theory. For detail, we refer readers to Jacobson's review \cite{Jacobson}, and \cite{Lin} for recent development. 
In  {\ae}-theory, the fundamental variables are   \cite{JM01},
\bq
\lb{2.0a}
\left(g_{\mu\nu}, u^{\mu}, \lambda\right),
\eq
where $g_{\mu\nu}$ is  the four-dimensional metric  of the space-time,  $u^{\mu}$  {is} the aether four-velocity, and $\lambda$ is a Lagrangian multiplier, which guarantees that the aether  four-velocity  is always timelike.
The general action of {\ae}-theory   is given by  \cite{Jacobson},
\bq
\lb{2.0}
S = S_{\ae} + S_{m},
\eq
where  $S_{m}$ denotes the action of matter,  and $S_{\ae}$  the gravitational action of the $\ae$-theory, given by
\bqn
\lb{2.1}
 S_{\ae} &=& \frac{1}{16\pi G_{\ae} }\int{\sqrt{- g} \; d^4x \Big[R(g_{\mu\nu}) + {\cal{L}}_{\ae}\left(g_{\mu\nu}, u^{\alpha}, {\lambda}\right)\Big]},\nb\\
S_{m} &=& \int{\sqrt{- g} \; d^4x \Big[{\cal{L}}_{m}\left(g_{\mu\nu}, u^{\alpha}; \psi\right)\Big]}.
\eqn
Here $\psi$ collectively denotes the matter fields, $R$    and $g$ are, respectively, the  Ricci scalar \footnote{Note that  $R$ here is different from the one used in Sec.II.A which denotes the distance.} and determinant of $g_{\mu\nu}$,
 and
\bq
\lb{2.2}
 {\cal{L}}_{\ae}  \equiv - M^{\alpha\beta}_{~~~~\mu\nu}\left(D_{\alpha}u^{\mu}\right) \left(D_{\beta}u^{\nu}\right) + \lambda \left(g_{\alpha\beta} u^{\alpha}u^{\beta} + 1\right),
\eq
 where $D_{\mu}$ denotes the covariant derivative with respect to $g_{\mu\nu}$, and  $M^{\alpha\beta}_{~~~~\mu\nu}$ is defined as
\bqn
\lb{2.3}
M^{\alpha\beta}_{~~~~\mu\nu} \equiv c_1 g^{\alpha\beta} g_{\mu\nu} + c_2 \delta^{\alpha}_{\mu}\delta^{\beta}_{\nu} +  c_3 \delta^{\alpha}_{\nu}\delta^{\beta}_{\mu} - c_4 u^{\alpha}u^{\beta} g_{\mu\nu}.\nb\\
\eqn
The four coupling constants $c_i$'s are all dimensionless, and $G_{\ae} $ is related to  the Newtonian constant $G_{N}$ via the relation \cite{CL04},
\bq
\lb{2.3a}
G_{N} = \frac{G_{\ae}}{1 - \frac{1}{2}c_{14}},
\eq
where $c_{ij}\equiv c_i +c_j$.  

It is easy to show that the Minkowski spacetime  is a solution of $\ae$-theory, in which the aether is aligned along the time direction, $\bar{u}_{\mu} = \delta^{0}_{\mu}$. 
 Then, the linear perturbations around the Minkowski background show that the theory in general possess three types of excitations, scalar, vector and tensor modes  \cite{JM04}, with their squared  {speeds given,  respectively, by}
\begin{eqnarray}
 \label{CSs}
 c_S^2 & = & \frac{c_{123}(2-c_{14})}{c_{14}(1-c_{13}) (2+c_{13} + 3c_2)}\,,\nonumber\\
 c_V^2 & = & \frac{2c_1 -c_{13} (2c_1-c_{13})}{2c_{14}(1-c_{13})}\,,\nonumber\\
 c_T^2 & = & \frac{1}{1-c_{13}},
\end{eqnarray}
where $c_{ijk} \equiv c_i + c_j + c_k$.  

Recently,   the combination of the gravitational wave event GW170817 \cite{GW170817},  and the event of the gamma-ray burst
GRB 170817A \cite{GRB170817} provides  a remarkably stringent constraint on the speed of the spin-2 mode, 
\bq
\lb{CD6}
- 3\times 10^{-15} < c_T -1 < 7\times 10^{-16},
\eq
which, together with Eq.(\ref{CSs}), implies that 
\bq
\lb{2.8a}
\left |c_{13}\right| < 10^{-15}.
 \eq
 Imposing further the following conditions: (a) the theory is free of ghosts; (b) the squared speeds $c_I^2\; (I = S, V, T)$ must be non-negative; (c) $c_{I}^2-1$ must be greater than $-10^{-15}$ or so, 
 in order to avoid the existence of the vacuum gravi-\v{C}erenkov radiation by matter such as cosmic rays \cite{EMS05}; and (d) the theory must be consistent with the current observations on the primordial 
 helium abundance $\left|G_{cos}/G_{N} - 1\right| \lesssim 1/8$, where $G_{cos} \equiv G_{\ae}/(1+ (c_{13} + 3c_2)/2)$ \cite{CL04}, together with Eq.(\ref{2.8a}) and the conditions,
\bq
\lb{CD5}
\left| \alpha_1\right| \le 10^{-4}, \quad 
 \left|\alpha_2\right| \le 10^{-7},
 \eq
from the  Solar System observations   \cite{Will06},  it was found that   the parameter space of the theory is restricted to \cite{OMW18}, 
 \bq
\lb{2.8b}
0 \lesssim c_{14} \lesssim c_{1}, \quad  c_{14} \lesssim c_2 \lesssim 0.095, \quad  c_{14} \lesssim 2.5\times 10^{-5}.
 \eq

Finally, we note that the theoretical and observational constraints of $\ae$-theory and gravitational waves were also studied in \cite{GHLP18}.


\begin{thebibliography}{nbound}

\bibitem{Einstein} E. Schucking {\em et al.}, The Collected Papers of Albert Einstein vol6, The Berlin Years, (Princeton University Press, Princeton, New Jersey, 1997). 

\bibitem{GW150914}  B.P. Abbott,   {\em et al.},  [LIGO Scientific and Virgo Collaborations] Phys. Rev. Lett. {\bf 116},  061102 (2016).  

\bibitem{GW151226}  B.P. Abbott,   {\em et al.},  [LIGO Scientific and Virgo Collaborations] Phys. Rev. Lett. {\bf 116},  241103 (2016).

\bibitem{GW170104}  B.P. Abbott,   {\em et al.},  [LIGO Scientific and Virgo Collaborations] Phys. Rev. Lett. {\bf 118},   221101 (2017).

\bibitem{GW170608}  B.P. Abbott,   {\em et al.},  [LIGO Scientific and Virgo Collaborations] Astrophys. J.  {\bf 851},  L35 (2017).

\bibitem{GW170814}  B.P. Abbott,   {\em et al.},  [LIGO Scientific and Virgo Collaborations] Phys. Rev. Lett. {\bf 119},  141101 (2017).

\bibitem{GW170817}  B.P. Abbott,   {\em et al.},  [LIGO Scientific and Virgo Collaborations] Phys. Rev. Lett. {\bf 119},  161101 (2017).

\bibitem{GWs} B.P. Abbott, et al. [LIGO/Virgo Collaborations], arXiv:1811.12907.

\bibitem{FC17} K. Fuhrmann,   {\em et al},  Astrophys. J. {\bf 836} 139 (2017) .

\bibitem{Tok06} A. Tokovinin,   {\em et al},  A$\&$A {\bf 450} 681 (2006) .

\bibitem{Ransom14} S.M. Ransom, {\em et al}, Nature {\bf 505}, 520 (2014) . 

 \bibitem{Shao16} L. Shao, Phys. Rev. D{\bf 93}, 084023 (2016).

\bibitem{Archibald18} A.M. Archibald, {\em et al}, Nature {\bf 559}, 73 (2018).

\bibitem{Jacobson} T. Jacobson, Einstein-aether gravity: a status report, arXiv:0801.1547.

\bibitem{Brans} C. Brans and R. H. Dicke, Mach's Principle and a Relativistic Theory of Gravitation, Phys. Rev. {\bf 124},  925 (1961).

\bibitem{AS16} A. Sesana, Prospects for Multi-band Gravitational-Wave Astronomy after GW150914, Phys. Rev. Lett. {\bf 116}, 231102 (2016). 

\bibitem{Foster07} B. Z. Foster, Phys. Rev. D{\bf 76}, 084033 (2007).
 
\bibitem{Yagi14}  K. Yagi,  D. Blas, E. Barausse, and N. Yunes,   Phys. Rev. D{\bf 89}, 084067 (2014).
 
\bibitem {Will18} C. Will, Class. Quantum Grav. {\bf 35}, 085001 (2018).

\bibitem{MM} M. Maggiore, Gravitational Waves Volume 1: Theory and Experiments (Oxford University Press, New York, 2016).

\bibitem{Moore15} C. J. Moore, R. H. Cole and C. P. L. Berry, Class. Quantum Grav. {\bf 32},  015014 (2015).

\bibitem{Lin} K. Lin, {\em et al.}, Gravitational wave forms, polarizations, response functions and energy losses of triple systems in Einstein-Aether theory, Phys. Rev. D{\bf 99}, 023010 (2019) [arXiv:1810.07707].

\bibitem{OMW18} J. Oost, S. Mukohyama and A. Wang, Constraints on Einstein-aether theory after GW170817, Phys. Rev. D{\bf 97},  124023 (2018) [arXiv:1802.04303].

\bibitem{Zhao19} X. Zhao,  {\em et al.}, in preparation (2019). 

\bibitem{Will} C.M. Will, Gravitational radiation, close binary systems, and the Brans-Dicke theory of gravity, Astrophys. J. {\bf 214}, 826 (1997).


\bibitem{Zhang17} X. Zhang, {\em et al.}, Phys. Rev. D{\bf 95}, 124008 (2017). 

\bibitem{Dmitra} V. Dmitrasinovic, M. Suvakov, and A. Hudomal, Phys. Rev. Lett. {\bf 113}, 101102 (2014). 

\bibitem{Foster} B.Z. Foster, Radiation Damping in Einstein-Aether Theory, arXiv:gr-qc/0602004v5.

\bibitem{Foster06} B.Z. Foster, Post-Newtonian parameters and constraints on Einstein-aether theory, Phys. Rev. D {\bf 73}, 064015 (2006).

\bibitem{BYY18} E. Berti, K. Yagi, and N. Yunes, Extreme gravity tests with gravitational waves from compact binary coalescences: (I) inspiral-merger, Gen. Relativ. Grav. {\bf 50}, 46 (2018). 

\bibitem{GRB170817} B. P. Abbott et. al.,  Virgo, Fermi-GBM, INTEGRAL, LIGO Scientific Collaboration, Gravitational Waves and Gamma-rays from a Binary Neutron Star Merger: GW170817 and GRB 170817A, Astrophys. J. {\bf 848} (2017) L13 [arXiv:1710.05834].

\bibitem{JM01} T.  Jacobson, D. Mattingly, Phys. Rev. D{\bf 64}, 024028 (2001).

\bibitem{CL04} S. M. Carroll and E. A. Lim, Phys. Rev. D{\bf 70}, 123525 (2004).

\bibitem{JM04} T.  Jacobson,    D. Mattingly, Phys. Rev. D{\bf 70}, 024003  (2004).

\bibitem{EMS05} J. W. Elliott, G. D. Moore and H. Stoica,  JHEP {\bf 0508}, 066 (2005) [arXiv:hep-ph/0505211].

\bibitem{Will06}  C. M. Will, Living Reviews in Relativity {\bf 9}, 3 (2006).

\bibitem{GHLP18}   Y.-G. Gong, S.-Q. Hou, D.-C. Liang, E. Papantonopoulos, Phys. Rev. D{\bf 97}, 084040 (2018).






\end{thebibliography}
\end{document}